\documentclass{emulateapj}
\usepackage{times}
\usepackage[colorlinks,citecolor=blue,linkcolor = blue]{hyperref}
\usepackage{url}
\usepackage{enumitem}  
\usepackage{graphics}
\usepackage{amsmath}
\usepackage{color}
\usepackage{float}
\setlist[enumerate]{wide=\parindent}
\usepackage{natbib}
\usepackage{subfigure}
\usepackage{appendix}
\usepackage{footnote}

\shorttitle{Expected Detection Rates of Field Dwarf Galaxies in the Local Volume}
\shortauthors{Danieli et al.}

\begin{document}

\title{Hunting Faint Dwarf Galaxies in the Field Using Integrated Light Surveys}

\author{Shany Danieli\altaffilmark{1,2,3}}
\author{Pieter van Dokkum\altaffilmark{3}}
\author{Charlie Conroy\altaffilmark{4}}

\altaffiltext{1}{Department of Physics, Yale University, New Haven, CT 06520, USA}
\altaffiltext{2}{Yale Center for Astronomy and Astrophysics, Yale University, New Haven, CT 06511, USA}
\altaffiltext{3}{Department of Astronomy, Yale University, New Haven, CT 06511, USA}
\altaffiltext{4}{Harvard-Smithsonian Center for Astrophysics, 60 Garden Street, Cambridge, MA, USA}

\begin{abstract}

We discuss the approach of searching low mass dwarf galaxies, $\lesssim10^6\textrm{ M}_{\odot}$, in the general field, using integrated light surveys. By exploring the limiting surface brightness-spatial resolution ($\mu_{\textrm{eff,lim}}-\theta$) parameter space, we suggest that faint field dwarfs in the Local Volume, between $3$ and $10 \textrm{ Mpc}$, are expected to be detected effectively and in large numbers using integrated light photometric surveys, complementary to the classical star counts method. We use a sample of Local Group dwarf galaxies to construct relations between their photometric and structural parameters, $\textrm{M}_{*}$-$\mu_{\textrm{eff,V}}$ and $\textrm{M}_{*}$-$\textrm{R}_{\textrm{eff}}$. We use these relations, along with assumed functional forms for the halo mass function and the stellar mass-halo mass relation, to calculate the lowest detectable stellar masses in the Local Volume and the expected number of galaxies as a function of the limiting surface brightness and spatial resolution. The number of detected galaxies depends mostly on the limiting surface brightness for distances  $>3\textrm{ Mpc}$ while spatial resolution starts to play a role at distances $>8\textrm{ Mpc}$. Surveys with $\mu_{\textrm{eff,lim}}\sim30\textrm{ mag arcsec}^{-2}$ should be able to detect galaxies with stellar masses down to $\sim10^4 \textrm{ M}_{\odot}$ in the Local Volume. Depending on the assumed stellar mass-halo mass relation, the expected number of galaxies between $3$ and $10\textrm{ Mpc}$ is $0.04-0.35\textrm{ deg}^{-2}$, assuming a limiting surface brightness of $\sim29-30\textrm{ mag arcsec}^{-2}$ and a spatial resolution $<4''$. We currently look for field dwarf galaxies by performing a blank wide-field survey with the Dragonfly Telephoto Array, optimized for the detection of ultra-low surface brightness structures.

\end{abstract}

\keywords{galaxies: abundances --- galaxies: dwarf --- galaxies: luminosity function, mass function}


\section{Introduction}\label{intro}

\begin{figure*}[t!]
{\centering
  \includegraphics[width=180mm]{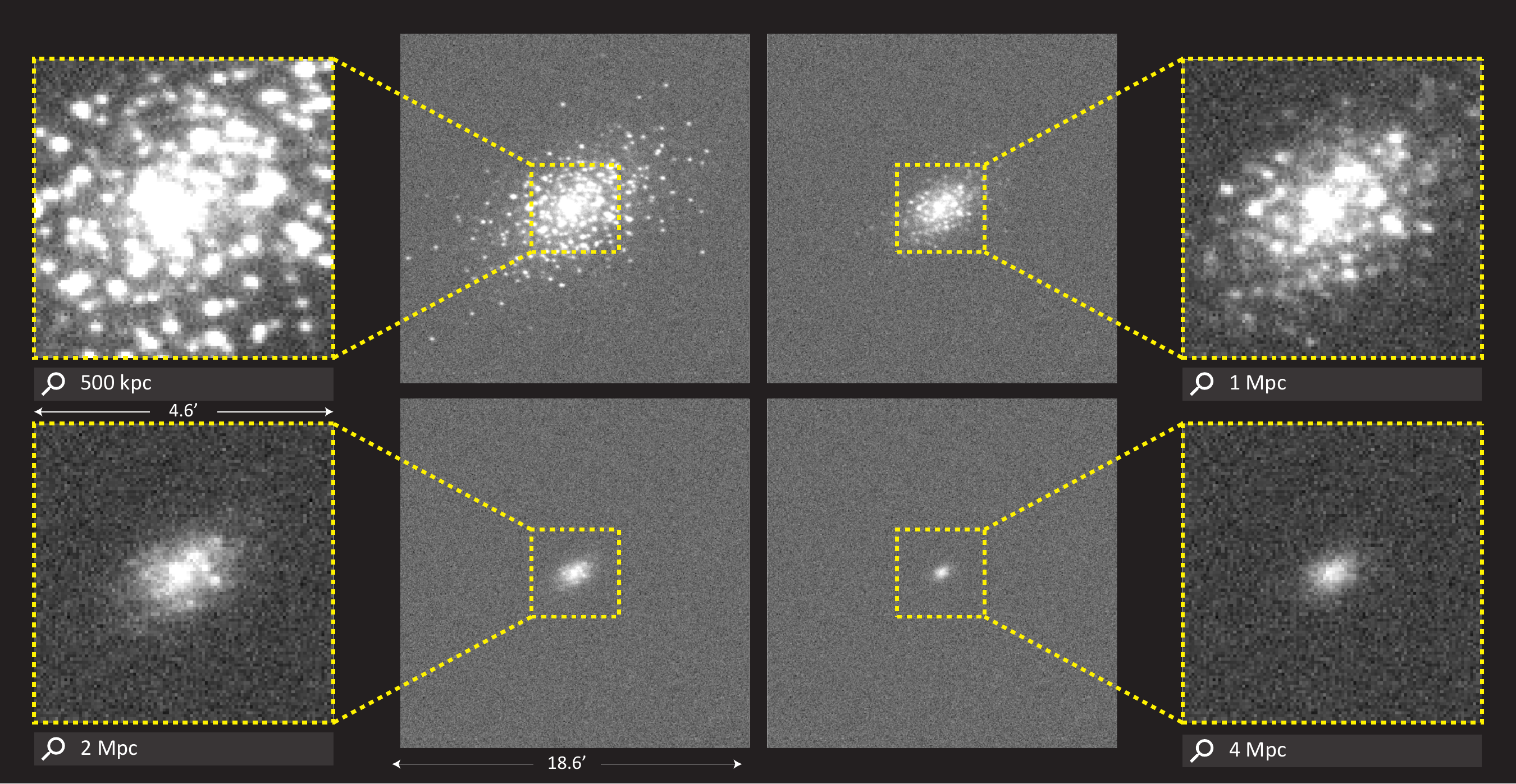}
  \caption{A dwarf galaxy with $M_* = 10^5 M_{\odot}$ as a function of increasing distance, artificially created using the \texttt{ArtPop} code. From a galaxy well-resolved into stars at $500 \textrm{ kpc}$, it gradually becomes less resolved and turns into an integrated flux object at $4 \textrm{ Mpc}$. The surface brightness remains constant while the size of the galaxy decreases with the distance.}
  \label{fig:fake_gal}
  }
\end{figure*}

The number of low mass dwarf galaxies in the Local Volume provides strong constraints on modern theories of galaxy formation (\citealt{2015MNRAS.454.1798K}).
There are currently no strong constraints on the lower mass cutoff of the luminosity function of galaxies. Of particular interest is whether the luminosity function extends all the way to Ultra Faint Dwarf (UFD) regime (e.g. \citealt{2009ApJ...692.1464G}) or there is a cutoff at higher masses. 
Related important and extensively discussed uncertainty regards the shape of the stellar mass-halo mass (SMHM) relation on the low mass end. There is no observational data for testing the abundance-matching-derived SMHM relation at low stellar masses (e.g. \citealt{2010MNRAS.404.1111G}; \citealt{2013ApJ...770...57B}; \citealt{2013MNRAS.428.3121M}; \citealt{2017MNRAS.465.4703K}) and further more, recent studies have shown a possible large scatter in this relation at low masses (\citealt{2017MNRAS.464.3108G}; \citealt{2017arXiv170506286M}). 

Therefore, performing a systematic deep, wide-field search for faint objects in the general field is of great importance.
This is also critical for our understanding of physical processes involved in low mass galaxy formation in the field. \citealt{2012ApJ...757...85G} showed that dwarf galaxies ($10^7<M_{*}<10^9 M_{\odot}$) with no active star formation are extremely rare ($<0.06\%$) in the field. It is interesting to examine this finding in the Local Volume, including further lower mass dwarfs. 

In the Local Group, dwarf galaxies have very low surface brightness, of $\gtrsim 28 \textrm{ mag arcsec}^{-2}$ (\citealt{2012AJ....144....4M}; \citealt{2015MNRAS.454.1798K}).
Many of the dwarf galaxies known to us today were at first detected using direct star counts. 
In this approach, galaxies are detected by identifying a stellar overdensity through individual star counts (compared to the density at larger scales, in order to confirm they are not part of a galactic background or foreground).
Star count surveys have proven to be very successful, with the detection of dozens of dwarf galaxies and star clusters in the Local Group and even slightly beyond (e.g. \citealt{1994ESOC...49...27I}; \citealt{2007ApJ...671.1591I}; \citealt{2008ApJ...686..279K}; \citealt{2009AJ....137..450W}; \citealt{2010ApJ...712L.103B}; \citealt{2011ApJ...732...76R}; \citealt{2013ApJ...772...15M}; \citealt{2015ApJ...805..130K}; \citealt{2015ApJ...807...50B}; \citealt{2016MNRAS.459.2370T}; \citealt{2016MNRAS.463..712T}). 
Studies based on star counts are able to reach effective surface brightness levels of $30 \textrm{ mag arcsec}^{-2}$ or fainter but suffer from another limiting factor. The brightness of stars, observed as point sources, decreases with the square of the distance and hence star count surveys are only efficient at identifying dwarf galaxies in the local universe. 

Beyond $5 \textrm{ Mpc}$, galaxies are more easily detected as integrated light objects, as surface brightness is independent of distance (for low redshift objects that do not suffer from cosmological surface brightness dimming of the form $(1+z)^{-4}$). In fact, many of the faint galaxies discovered in the Local Volume in recent years were detected as integrated light objects and have remarkably expanded the census of faint and ultra faint dwarf candidates beyond the Local Group (\citealt{2013AJ....145..101K}; \citealt{2014arXiv1401.2719K}; \citealt{2015AstBu..70..379K}; \citealt{2014ApJ...787L..37M}; \citealt{2016MNRAS.457L.103R}; \citealt{2016A&A...588A..89J}; \citealt{2017A&A...603A..18H}; \citealt{2017arXiv170103681M}). 
Furthermore, many recent surveys find faint, unresolved low surface brightness dwarf galaxies in nearby groups and clusters, demonstrating a possible gold mine for finding faint galaxies in the field (\citealt{2014ApJ...787L..37M}; \citealt{2016ApJ...824...10F}; \citealt{2017arXiv170103681M}; \citealt{2017arXiv170103681M}; \citealt{2017ApJ...847....4G}; \citealt{2017arXiv170904474G}).
Integrated light surveys offer complementary benefits and drawbacks compared to star count surveys: they are able to efficiently probe large volumes, but require extreme surface brightness sensitivity. Moreover, follow-up observations are required in order to measure distances and determine whether a galaxy is associated with a group of galaxies or is in its foreground or background (\citealt{2016ApJ...833..168M}; \citealt{2017ApJ...837..136D}).

Recent advances allow imaging large areas of the galactic sky down to ultra low surface brightness levels and provide an excellent platform for hunting dwarf galaxies in the field.
The tremendous progress in improving the surface brightness limit in the last few years was achievable by using a new innovative design that minimizes systematic errors that often limit the accuracy of background estimation and flat-fielding. 
The Dragonfly Telephoto Array (hereafter Dragonfly) is an example for such an imaging system. It was designed to overcome the systematic limitations that prevent conventional telescopes from being able to image down to low surface brightness levels (\citealt{2014PASP..126...55A}).
It is comprised of 48 high-end commercial telephoto lenses that feature nano-fabricated coatings with sub-wavelength structures to yield a factor of ten improvement in wide-angle scattered light relative to other conventional astronomical telescopes. Its performance is equivalent to that of a 1 meter aperture refractor with a f/0.39 focal ratio and a wide  field of view of six square degrees.
Dragonfly is specialized to efficiently observe extended objects down to hitherto unprecedentedly low surface brightness levels, and is therefore ideally suited to detect possible dwarfs candidates in the field. 
Dragonfly has already proven successful in identifying dozens of low surface brightness ($26-28 \textrm{ mag arcsec}^{-2}$) objects in various fields (\citealt{2014ApJ...787L..37M}; Cohen et al. in prep, Danieli et al. in prep).

In this paper, we demonstrate the importance of surface brightness as a key parameter in such systematic searches.
We also discuss the trade-off between surface brightness and resolution and what roles they both play in our ability to detect faint objects.  
We build on known cosmological models and on the census of dwarf galaxies in the Local Group to estimate the expected number of dwarf galaxies in the field.

This paper is organized as follows: we start by presenting a new tool, \texttt{ArtPop}, for simulating the appearance of galaxies in various photometric systems, using artificial stellar populations, in Section \ref{artpop}. We use \texttt{ArtPop} to demonstrate the variation in visibility of dwarf galaxies at different distances. 
Next, we explore the detectability of the lowest mass galaxies in the Local Volume (out to $10 \textrm{ Mpc}$) using integrated light, in Section \ref{integrated}.
We present a model for the expected number of field dwarfs in Section \ref{methods} and present our results in Section \ref{results}.
We then discuss the advantages of integrated light surveys compared to (ground-based) star counts surveys in certain regimes in Section \ref{vs}.
We conclude and discuss a systematic search for field dwarf galaxies, over a wide field in the Local Volume, using the Dragonfly Telephoto Array in Section \ref{discussion}.

\begin{figure*}[t]
\centering
  \includegraphics[width=\textwidth]{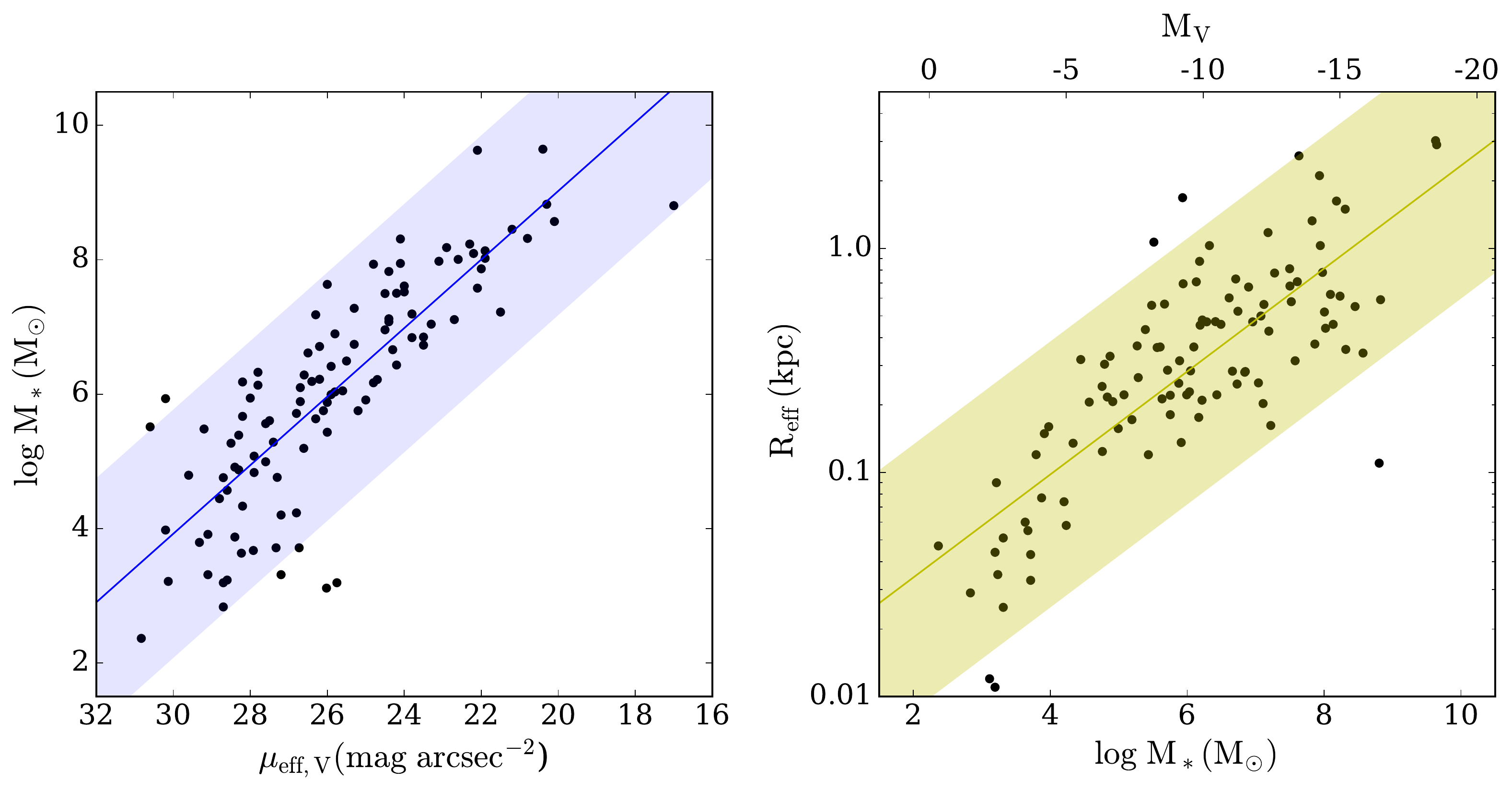}
  \caption{The stellar mass-effective surface brightness and effective radius-stellar mass relations of dwarf galaxies in the Local Group (\citealt{2012AJ....144....4M}; \citealt{2015ApJ...807...50B}; \citealt{2015ApJ...813..109D}; \citealt{2016MNRAS.459.2370T}; \citealt{2016MNRAS.463..712T}; \citealt{2016ApJ...832...21H}; \citealt{2017ApJ...838...11S}; \citealt{2017arXiv170405977H}), along with the best-fit of linear model for this data. Shaded regions show the $95\%$ confidence interval. Stellar masses have been estimated assuming a $V-$band mass-to-light ratio of $M/L_{V} = \textrm{2.0}$ (\citealt{2009ApJ...699..486C}).}
  \label{fig:relations}
\end{figure*}


\section{Simulated Images of Galaxies with \texttt{ArtPop}}\label{artpop}

In order to demonstrate how an ultra faint dwarf galaxy will be observed at different distances, we introduce a new tool, \texttt{ArtPop}, for simulating the appearance of galaxies in different photometric systems, using \texttt{Art}ificial stellar \texttt{Pop}ulations.
\texttt{ArtPop} requires three sets of parameters as input:

\begin{enumerate}
  \item The galaxy stellar population parameters: Initial Mass Function (IMF), stellar mass or number of stars ($M_{*}$ or $N_{\textrm{stars}}$), age and chemical composition (currently parameterized by [Fe/H]).
  \item Structural and spatial information: distance (D), effective (half-light) radius ($r_{\textrm{eff}}$), S\'ersic index (n), ellipticity ($\epsilon$) and position angle ($\theta$).
  \item Imaging and: photometric system (including magnitude zeropoint) and a Point Spread Function (PSF).
\end{enumerate}

\texttt{ArtPop} creates the artificial galaxy from stars in the following way: $N_{\textrm{stars}}$ stars are sampled according to the IMF from MIST isochrones (\citealt{2016ApJS..222....8D}; \citealt{2016ApJ...823..102C}). The selected stars are shifted to the right distance and distributed spatially according to a S\'ersic profile that serves as a probability distribution for their position. Then, each point source is convolved with the PSF of the imaging system.

In Figure \ref{fig:fake_gal} we show an \texttt{ArtPop} galaxy as observed with the Dragonfly Telephoto Array photometric system: SDSS $g$-band, $2."8 \textrm{ pixel}^{-1}$ and the Dragonfly PSF (\citealt{2014ApJ...787L..37M}).
The galaxy was constructed using $5 \cdot 10^{5} $ stars, corresponding to a stellar mass of $M_{*} = 10^{5} M_{\odot}$ for a Salpeter (\citeyear{1955ApJ...121..161S}) IMF, and has an effective radius of $400 \textrm{ pc}$.
At a distance of $500 \textrm{ kpc}$ (top left) the galaxy is easily resolved into stars and different stellar populations can be detected and quantified. However, at farther distances, the resolved galaxy transforms to appear as a low surface brightness ``blob"; less and less individual stars can be identified and the galaxy turns into a smooth object. 
Once its angular size is small than the spatial resolution of the imaging system, it will look like a point source and the brightness of the centered pixel will decrease as $\sim \textrm{D}^2$.
As can be seen from this example, a very low mass galaxy can be identified as a dwarf galaxy candidate at larger distances than those achieved in star counts surveys. 
In the next section we present a model for the expected number of field dwarf galaxies and a calculation for the required observational capabilities for testing these predictions.


\section{Discovering Dwarf Galaxy Candidates using Integrated Light Imaging}\label{integrated}

In this section we present a model (\autoref{methods}) and results (\autoref{results}) for calculating the abundances of dwarf galaxies in the field, between $3$ and $10 \textrm{ Mpc}$, using integrated light imaging.

\subsection{Methodology}\label{methods}

We start with compiling all known dwarf galaxies in the Local Group along with their photometric and structural parameters (\citealt{2012AJ....144....4M}; \citealt{2015ApJ...807...50B}; \citealt{2015ApJ...813..109D}; \citealt{2016MNRAS.459.2370T}; \citealt{2016MNRAS.463..712T}; \citealt{2016ApJ...832...21H}; \citealt{2017ApJ...838...11S}; \citealt{2017arXiv170405977H}). 
We use this observational data set to get an estimate for the number of dwarf galaxies in the field. An important assumption in our model is that dwarf galaxies in the field have similar statistical properties as dwarf satellite galaxies in the Local Group. Of course, field dwarfs might have different properties than dwarf galaxies in the Local Group due to cosmic variance and environmental effects that can impact their formation mechanisms (see, e.g., \citealt{2009ApJ...692.1464G}).
However, this is by far the most complete sample of dwarf galaxies down to the lowest masses, available to us.

We use the effective radii, magnitudes in $V$-band and ellipticities to calculate the mean $V$-band surface brightness within the effective radius, $\mu_{\mathrm{eff,}V}$. 
We assume a $V$-band mass-to-light ratio of $M/L_{V} = \textrm{2.0}$, appropriate for old (10 Gyr) metal-poor ([Z/H]  $<-1$) populations (\citealt{2009ApJ...699..486C}). The assumption of an old, metal-poor stellar population is conservative, as field dwarfs might have younger stellar populations, possibly even with active star formation (\citealt{2012ApJ...757...85G}).
We use this observational data set to obtain a relation between the stellar mass and the effective surface brightness and between the effective radius and the stellar mass.
The best-fit to the observational data is given by

\begin{equation}\label{eq:1}
	\log M_{*} = -0.51\cdot \mu_{\mathrm{eff},V}+19.23,
\end{equation}

\noindent with a 2$\sigma$ scatter of 0.92 dex, and 

\begin{equation}\label{eq:2}
	\log R_{\mathrm{eff}} = 0.23\cdot \log M_{*}-1.93,
\end{equation}

\noindent with a 2$\sigma$ scatter of 0.29 dex.

The compiled data along with the best-fit relations are presented in Figure \ref{fig:relations}.

\begin{figure}[t]
	\includegraphics[width=0.49\textwidth]{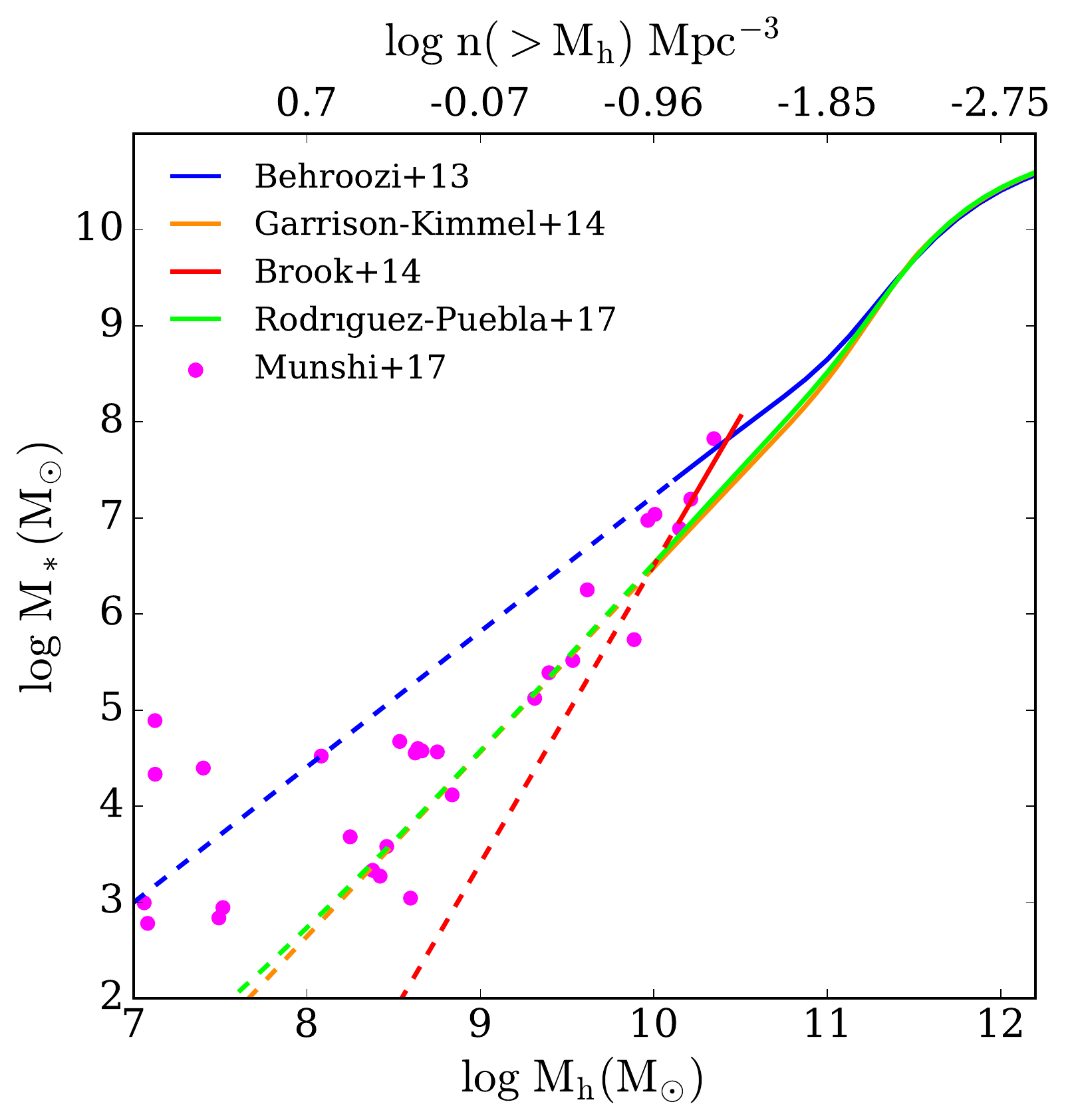}
  	\caption{Various published stellar mass-halo mass relations (\citealt{2013ApJ...770...57B}; \citealt{2014MNRAS.438.2578G}; \citealt{2014ApJ...784L..14B}; \citealt{2017arXiv170304542R}). Dotted lines are extrapolations of derived relations and the pink circles are galaxies from the 40 Thieves simulation (\citealt{2017arXiv170506286M}). The upper $x-$axis has been converted to halo cumulative number density based on our fiducial cosmology.} 
  	\label{fig:smhm}
\end{figure}

\begin{figure*}[t]
\centering
  \includegraphics[width=\textwidth]{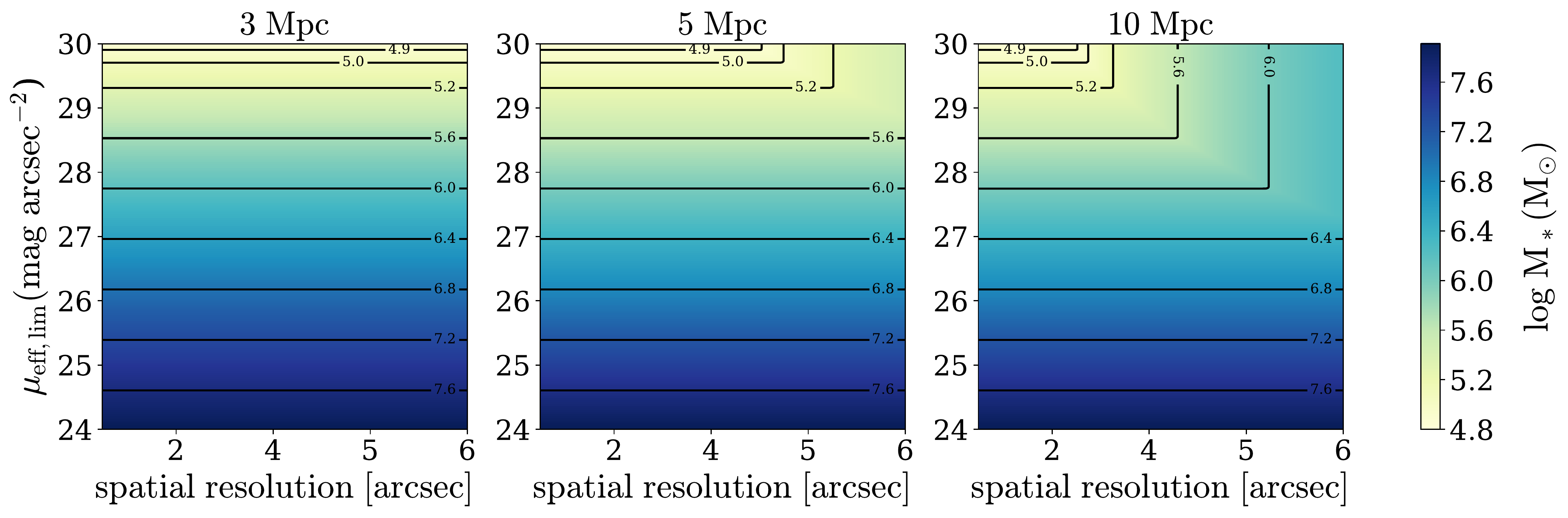}
  \caption{The minimal detectable stellar mass in integrated light surveys, assuming a limiting effective surface brightness in the $V-$band and a spatial resolution. Black contour lines indicate constant minimal stellar mass for different values of limiting effective surface brightness and spatial resolution. At the indicated masses, the completeness is $95 \%$. }
  \label{fig:min_stellar_mass}
\end{figure*}

Another key ingredient in our calculation is the stellar mass-halo mass relation for galaxies, $\textrm{M}_{*}-\textrm{M}_{h}$.
A powerful and widely used technique to derive this relation is by using the abundance matching ansatz. In its simplest implementation, observed galaxies are matched in a one-to-one fashion with dark matter halos from a dark matter-only simulation while assuming a monotonic relation between the stellar mass, $M_{*}$ and the dark matter halo mass, $M_h$, such that the cumulative number density of dark matter halos matches the cumulative number density of galaxies (e.g. \citealt{1988ApJ...327..507F}; \citealt{2003MNRAS.339.1057Y}; \citealt{2004ApJ...609...35K}; \citealt{2006ApJ...647..201C}; \citealt{2006MNRAS.371.1173V}; \citealt{2006MNRAS.371.1173V}; \citealt{2010MNRAS.404.1111G}; \citealt{2013ApJ...770...57B}; \citealt{2013MNRAS.428.3121M}; \citealt{2014ApJ...784L..14B}; \citealt{2014MNRAS.438.2578G}; \citealt{2015MNRAS.448.2941S}; \citealt{2017arXiv170304542R}; \citealt{2017arXiv170506286M}; \citealt{2017MNRAS.467.2019R}; \citealt{2017arXiv170505373M}).
Although abundance matching studies are in good agreement for halos of masses $\gtrsim 10^{11} \textrm{M}_{\odot}$, at lower masses there is a large uncertainty in the stellar mass-halo mass relation. Different studies present various slopes for the relation below stellar masses of a few $\times 10^7 \textrm{M}_{\odot}$, presumably due to incompleteness at the low mass end and due to the variations in the halo mass function assumed in various simulations. 
The derived slopes of the low mass $\textrm{M}_{*}-\textrm{M}_{h}$ relation, $\alpha$, where ${M}_{*} \propto M_h^{\alpha}$, span a wide range of $1.6-3.1$.
Moreover, while the scatter at the high mass end is consistently measured to be relatively small, $\sim$ 0.2 dex or less, low luminosity galaxies have a more stochastic star formation, resulting in a large scatter.

Recent studies have explored the significance of the scatter in the $\textrm{M}_{*}-\textrm{M}_{h}$ relation and quantified the scatter in this relation for low mass galaxies (\citealt{2017MNRAS.464.3108G}; \citealt{2017arXiv170506286M}; \citealt{2018MNRAS.473.2060J}). 
The uncertainty in the  $\textrm{M}_{*}-\textrm{M}_{h}$ relation at low masses can have a large effect on our predictions. 
In this calculation we adopt two relations - the relation from Rodriguez-Puebla et al. \citeyear{2017arXiv170304542R} as a lower limit and the relation from Behroozi et al. \citeyear{2013ApJ...770...57B} as an upper limit. 
In Figure \ref{fig:smhm} we show these two relations along with other recently derived stellar mass-halo mass relations.
We calculate the cumulative halo number density as a function of halo mass assuming a dark matter halo mass function from \citealt{2010ApJ...724..878T}, obtained using the \texttt{HMFcalc} code (\citealt{2013A&C.....3...23M}). We adopt cosmological parameters consistent with the 7 year WMAP results (\citealt{2011ApJS..192...18K}): $H_0=70.4 \textrm{ km s}^{-1}$, $\Omega_{m}=0.27=1-\Omega_{\Lambda}$ and $\sigma_8=0.81$.
The upper $x$-axis of Figure \ref{fig:smhm} shows the values of the cumulative number density as a function of halo mass.

Given the model ingredients described above, we can calculate the expected number of dwarf galaxies with a particular distance, size, and surface brightness. The number of \emph{detected} galaxies also depends on the imaging capabilities, in particular, the limiting surface brightness, $\mu_{\mathrm{eff,lim}}$, and the spatial resolution, $\theta$. 

For a given limiting surface brightness, $\mu_{\mathrm{eff,lim}}$, we use the linear relation shown in equation \ref{eq:1} to get the estimated value for the limiting stellar mass, $M_{*,\mathrm{lim}}$. In order to keep our calculation conservative we consider the value of the stellar mass after adding a 2$\sigma$ scatter, i.e., we use: $\log M_{*,\mathrm{lim}} = -0.51\cdot \mu_{\mathrm{eff},\mathrm{lim}}+19.23 + 2\sigma_{\log M_{*}}$, where $\sigma$ is the standard deviation. We then use the relation shown in equation \ref{eq:2} to estimate the effective radius of the galaxies with such stellar masses, considering the smallest detectable objects to be $\log R_{\mathrm{eff,lim}} = 0.23\cdot \log M_{*}-1.93 -2\sigma_{\log R_{\mathrm{eff}}}$, i.e., within 2$\sigma$ range of the average effective radius. 

In the next step we consider the spatial resolution in arcseconds, $\theta$, which determines the limiting physical size of detectable objects and thus the visible horizon for the lowest detectable stellar masses.
Given a spatial resolution, $\theta$, objects with limiting effective radius $R_{\mathrm{eff,lim}}$ can be identified as galaxies to distances of
\begin{equation}\label{eq:3}
	D_{\mathrm{lim}}(\mu_{\mathrm{eff,lim}},\theta) = \frac{2.06\cdot10^{5} \cdot R_{\mathrm{eff,lim}}\textrm{[pc]}}{\theta \textrm{ [arcsec]}},
\end{equation}

with effective radius calculated as described above: 
\begin{equation}\label{eq:4}
\begin{aligned}
	\log(R_{\mathrm{eff,lim}}) = 0.23\cdot [-0.51\cdot \mu_{\mathrm{eff,lim}}+19.23 + 2\sigma_{\log M_{*}}] \\
	-1.93 -2\sigma_{\log R_{\mathrm{eff}}},
\end{aligned}	
\end{equation}

where $\sigma_{\log M_{*}} = 0.92$ and $\sigma_{\log R_{\mathrm{eff}}} = 0.29 $. 

The described model ingredients are combined to calculate the predicted cumulative number of galaxies, as a function of stellar mass, limiting surface brightness and resolution, in the Local Volume ($D_{\mathrm{LV}}$=10 Mpc), in the following way:
\begin{widetext}
\begin{equation}\label{eq:6}
	N(>M_{\mathrm{*,lim}}(\mu_{\mathrm{eff,lim}}),\theta) = \frac{4\pi}{3 \cdot 41253 } \begin{cases}
		 \int_{M_{\mathrm{*,lim}}}^{M_*(\theta,D=10)} \frac{dn}{dM} D_{\textrm{lim}}^3((M,\theta))dM + n(>M_*(\theta,D=10))D_{\mathrm{LV}}^3 , & \mathrm{if } \quad D_{\mathrm{lim}} < 10 \\
		n(>M_*(\theta,D=10))D_{\mathrm{LV}}^3 , & \mathrm{if } \quad D_{\mathrm{lim}} \geqslant 10 
	\end{cases}
\end{equation}
\end{widetext}

where $\frac{dn}{dM}$ is the differential stellar mass function, converted from the differential halo mass function using the $\textrm{M}_{*}-\textrm{M}_{h}$ relation. 
The lower limit of the integration is given by $M_{*,\mathrm{lim}}$, and the upper limit is the lowest stellar mass that can be detected out to the edge of the Local Volume, $M_*(\theta,D=10)$. $n(>M_*(\theta,D=10))$ is the cumulative number density of galaxies for stellar mass larger than $M_*(\theta,D=10)$, again converted from the cumulative number density of halos using the $\textrm{M}_{*}-\textrm{M}_{h}$ relation. 

In the next section we present the results we obtain using the model just described.

\subsection{Results}\label{results}

\begin{figure*}
\centering
\subfigure
{
   \includegraphics[width=\textwidth]{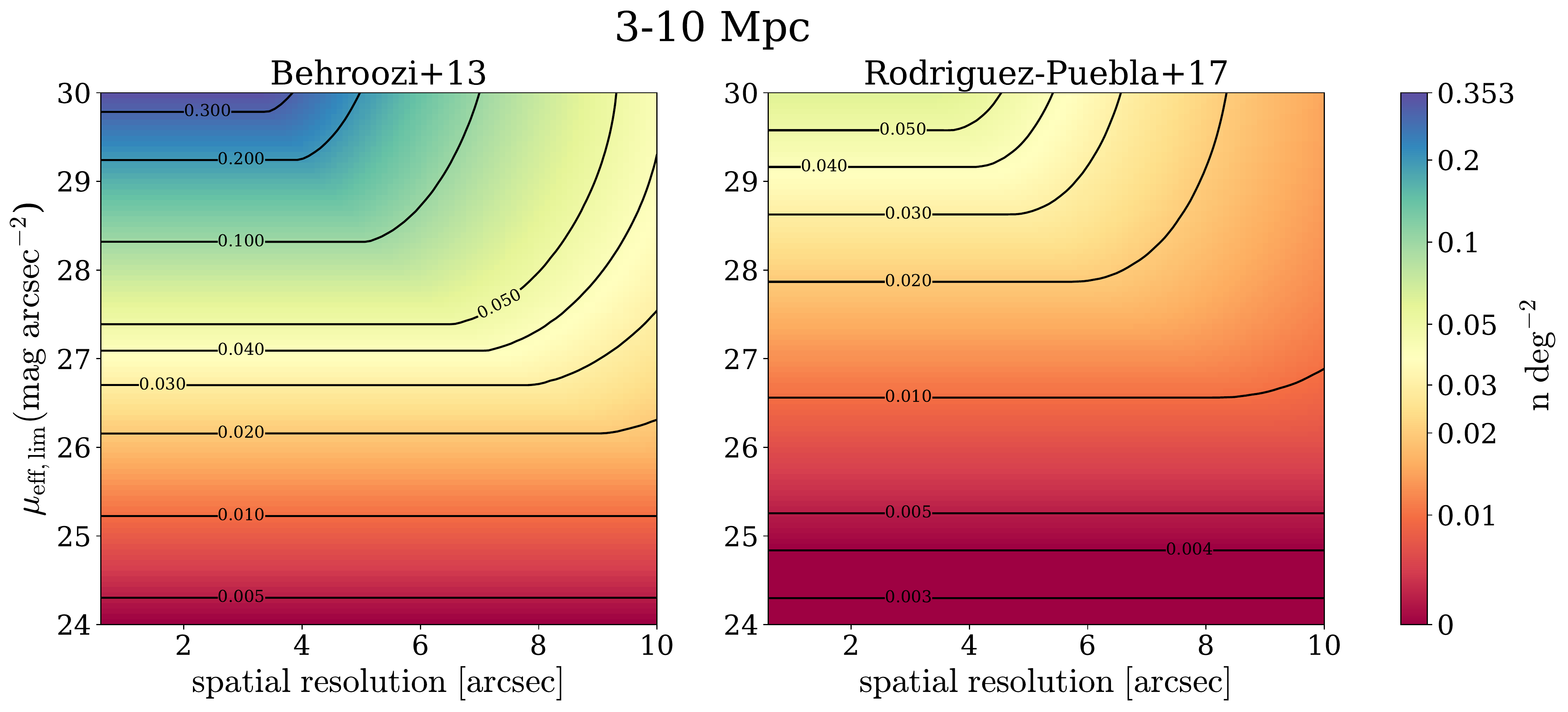}
   \label{fig:Cum1} 
}

\subfigure
{
   \includegraphics[width=\textwidth]{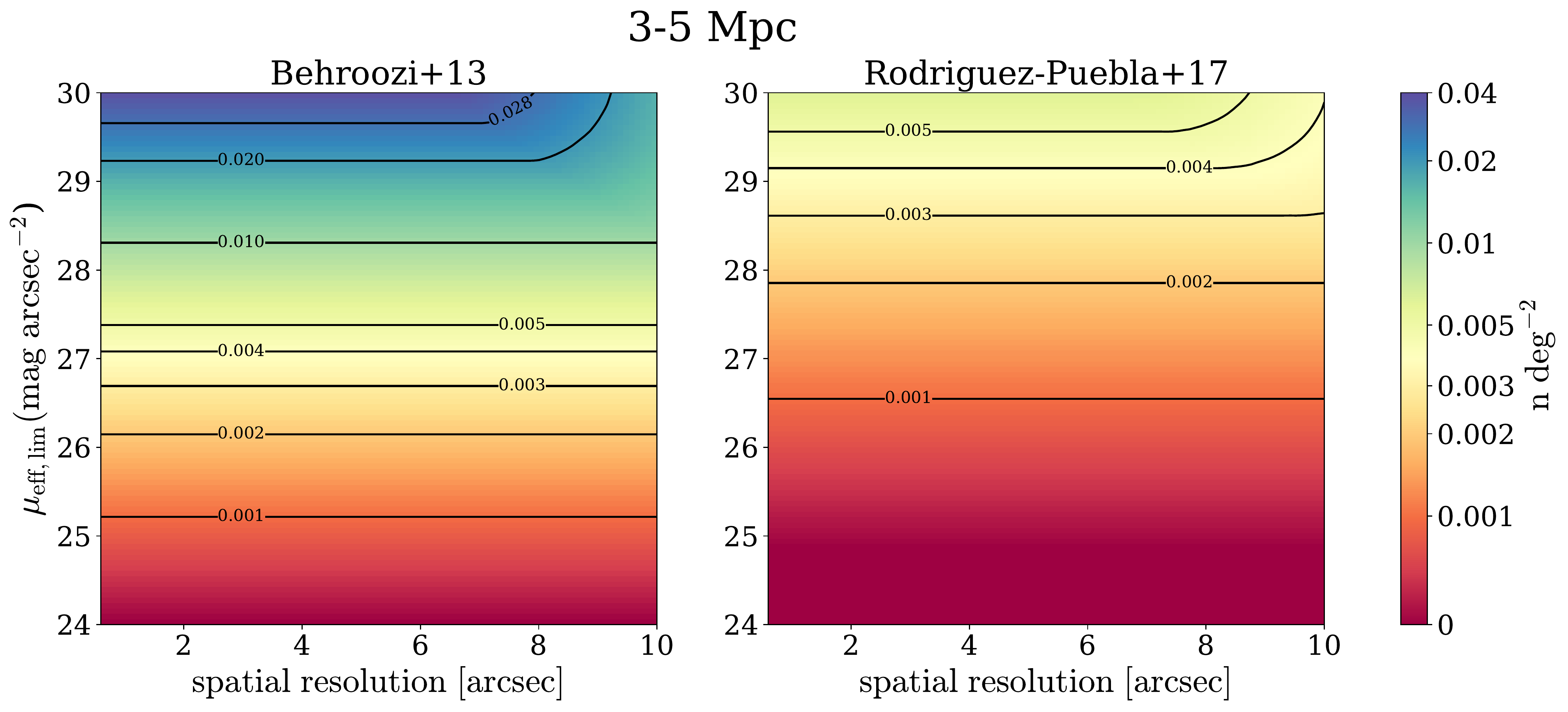}
   \label{fig:Cum2}
}
\caption{The cumulative number of predicted field galaxies per square degree to be detected in the Local Volume, between $3$ and $10$ Mpc (upper panel) and between $3$ and $5$ Mpc (lower panel) using an integrated light imaging, assuming a limiting effective surface brightness in $V$-band, $\mu_\textrm{eff,lim}$, and a spatial resolution, $\theta$. The left and right panels were calculated assuming the Behroozi et al. (\citeyear{2013ApJ...770...57B}) and the Rodrigues-Puebla et al. (\citeyear{2017arXiv170304542R}) stellar mass-halo mass relations, respectively.}
\label{fig:abundances}
\end{figure*}

\subsubsection{Detection Limits}\label{limits}

As described in Section \ref{methods}, assuming a limiting surface brightness and spatial resolution we use the linear relations, $\mu_{\textrm{eff},V}-\log M_{*}$ and $\log M_{*}-\log R_{\textrm{eff}}$, that were constructed using the empirical measured properties of dwarf galaxies in the complete Local Group sample, to infer the minimal detectable stellar mass for a set of ($\mu_{\textrm{eff}}$,$\theta$) at various distances. 
In Figure \ref{fig:min_stellar_mass}, we show the minimal detectable stellar mass  as a function of limiting surface brightness and spatial resolution, for different distances.
For all distances, the minimal detectable stellar mass depends strongly on the limiting surface brightness, which is a crucial parameter when carrying our integrated light surveys for the purposes of detecting new low mass galaxies. 
The limiting stellar mass changes quite dramatically ranging from stellar masses of $\sim 10^8 M_{\odot}$ for surveys with limiting surface brightness of $\sim 24 \textrm{ mag arcsec}^{-2}$ to stellar masses as low as $\sim 10^4 M_{\odot}$ for limiting surface brightness of $\sim 30 \textrm{ mag arcsec}^{-2}$. 
While the surface brightness limit plays such an important role for the entire range of distances examined, spatial resolution starts to impact around distances of $8 \textrm{ Mpc}$.
Detection at the outskirts of the Local Group, at $3 \textrm{ Mpc}$, are entirely independent of spatial resolution over the range that we probed.
Limited to the Local Volume, even with a low spatial resolution of $\theta \sim 5 \textrm{ arcsec}$, extremely low mass galaxies are potentially detectable. 

In Figure \ref{fig:min_stellar_mass} we show mass limits for $95 \%$ completeness. In the following we include all galaxies that fall within the ($\theta,\mu$) limits.
Clearly, the minimal detectable stellar mass shown here affects the number of predicted field galaxies in the Local Volume, presented in the next section. 

\subsubsection{Detection Rates of Field Dwarf Galaxies in the Local Volume}\label{abundance}

The resulting model predicted detection rates of dwarf galaxies in the field for two volumes, $3-10 \textrm{ Mpc}$ and $3-5 \textrm{ Mpc}$, are shown in Figure \ref{fig:abundances}.
We show results for two stellar mass-halo mass relations, Rodrigues-Puebla et al. (\citeyear{2017arXiv170304542R}, hereafter, RP17) in the right panel and Behroozi et al. (\citeyear{2013ApJ...770...57B}, hereafter B13) in the left panel, in order to get lower and upper limits for the estimated values as well as to highlight the sensitivity of the adopted model and the importance of constraining this relation observationally at low masses. 
The figures show the expected number of galaxies per square degree depending on limiting surface brightness and spatial resolution. 

The expected number of detected field galaxies varies significantly given the limiting surface brightness and the spatial resolution as well as different stellar mass-halo mass relations, ranging between $0.002$ and and $0.35$ galaxies per square degree in the Local Volume (Figure \ref{fig:abundances}).
The largest number of galaxies is obtained when we adopt the B13 stellar mass-halo mass relation (left panels). There, the number of predicted detected galaxies is as high as 0.35 galaxies per square degree, for limiting surface brightness levels fainter than $\mu_{\textrm{eff},\textrm{lim}} \sim 29.5 \textrm{ mag arcsec}^{-2}$ and a spatial resolution better than $\theta \sim 3.5 "$.
However, adopting the RP17 relation reduces the number of detected field dwarfs for the same observational limits to $\sim 0.05$ galaxies per square degree. 
Consistently with the results presented in \ref{limits}, the limiting surface brightness plays an important role in the predicted number of galaxies while only at surface brightness limits of $\gtrsim 26.5 \textrm{ mag arcsec}^{-2}$ the spatial resolution starts to be important. Then, the number of predicted galaxies per square degree increases as we lower the limiting surface brightness and the spatial resolution, as expected. 

For the smaller volume, $3-5 \textrm{ Mpc}$, the predicted number is obviously much smaller, ranging between $0$ and and $0.04$ galaxies per square degree, depending on the surface brightness limit and the spatial resolution. 
Considering galaxies in this volume, the spatial resolution seems to be an almost insignificant parameter in the context of only detecting the galaxies. Of course, better spatial resolution is crucial for resolving the galaxies into their different stellar populations.  
Similar to the results for the larger volume, assuming the two stellar mass-halo mass relations, B13 and RP17, results in significantly different values for the predicted cumulative number.
Comparing the two panels in each figure, it is easy to notice that at fixed number density there is almost two orders of magnitudes difference in their brightness between RP17 and B13.
These remarkable differences when adopting two different stellar mass-halo mass relations emphasize the necessity of detecting these dwarfs and placing strong constraints the stellar mass-halo mass relation at low masses.


\section{Comparison to Star Counts}\label{vs}

\begin{figure*}[t]
	\centering
	\includegraphics[width=0.5\textwidth]{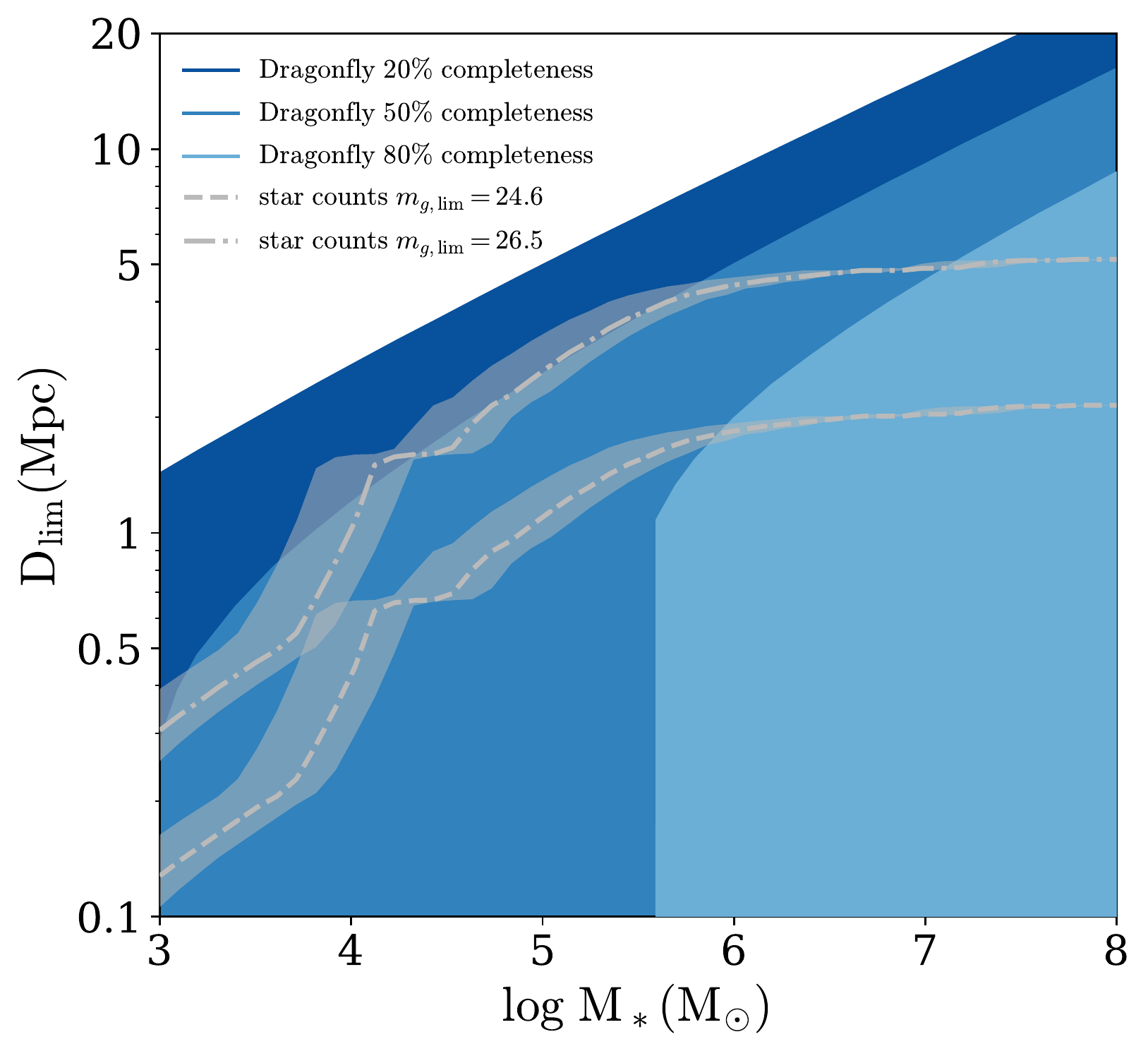}
  	\caption{Limiting distances at a function of stellar mass for integrated light imaging with the Dragonfly Telephoto Array ($\mu_{\textrm{eff,lim}}=29.5 \textrm{ mag arcsec}^{-2}$) and using the star counts method in surveys limited by magnitudes of $24.6$ and $26.5 \textrm{ mag}$ corresponding to the DECam and Hyper Supreme Cam Surveys. The blue shades show the limiting distances for galaxies that are detectable in $20$, $50$ and $80\%$ of a large sample of simulated galaxies. The limiting distances for the Dragonfly $20\%$ and $50\%$ completeness levels is mostly dominated by the spatial resolution, the $80\%$ sample limiting distance drops quickly at logarithmic stellar masses of $\sim 5.5 \textrm{ M}_{\odot}$ due to the limiting surface brightness.}
  	\label{fig:comparison}
\end{figure*}

\begin{figure*}[t]
\centering
  \includegraphics[width=\textwidth]{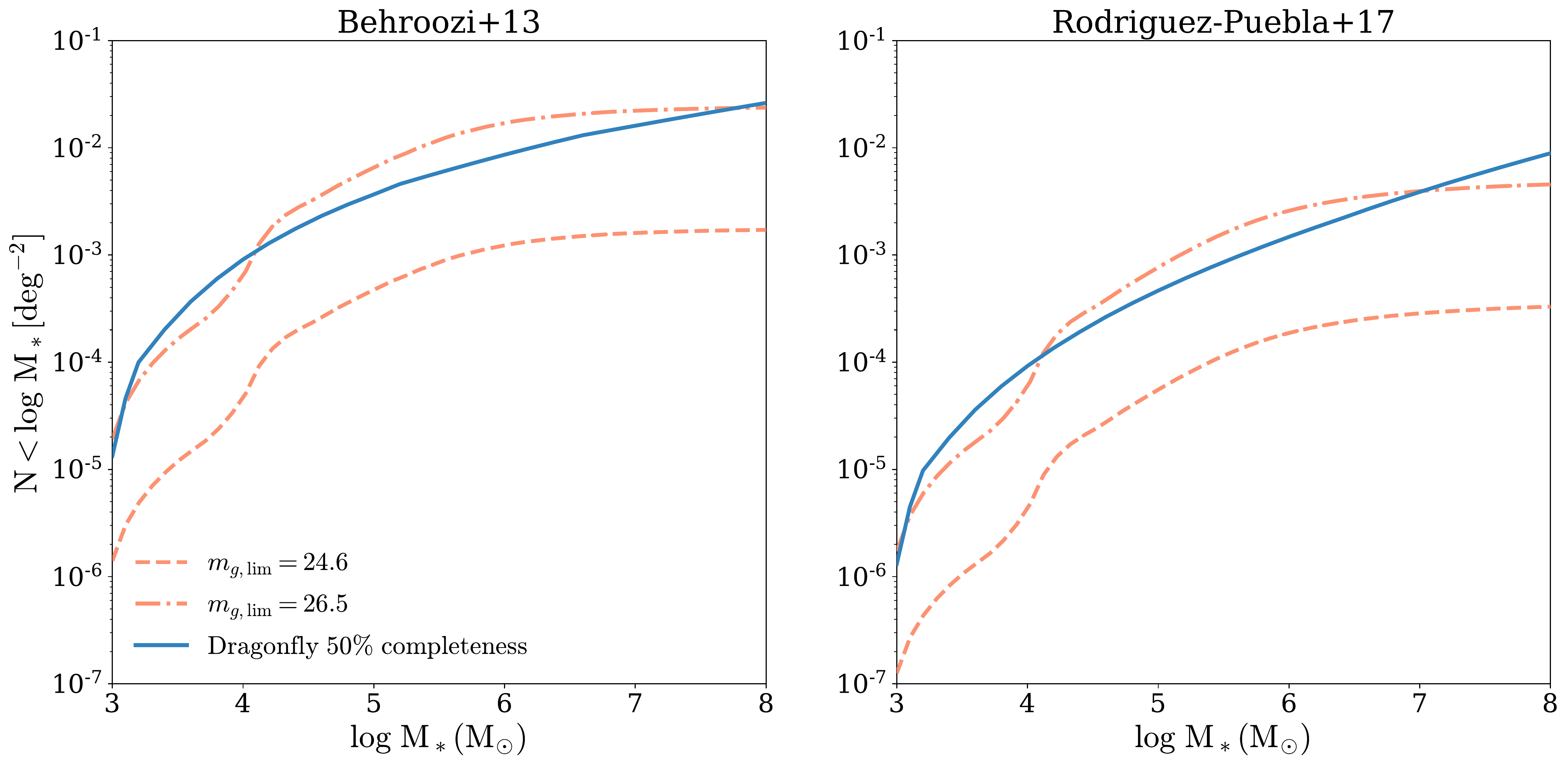}
  \caption{The cumulative number of the predicted detection rates of galaxies with the Dragonfly Telephoto Array (blue; $50\%$ completeness, see Figure \ref{fig:comparison}) and by using the star counts method in surveys limited by magnitudes of $24.6$ and $26.5 \textrm{ mag}$ corresponding to the DECam and Hyper Supreme Cam Surveys (dashed light pink curves). The left and right panels were calculated assuming the Behroozi et al. (\citeyear{2013ApJ...770...57B}) and the Rodrigues-Puebla et al. (\citeyear{2017arXiv170304542R}) stellar mass-halo mass relations, respectively.}
  \label{fig:cumulative}
\end{figure*}

After quantifying the detectability of dwarf galaxies using integrated light surveys, we now turn to compare our results to expectations from star count surveys.
For an integrated light survey we adopt the values from the Dragonfly Nearby Galaxies Survey (\citealt{2014ApJ...787L..37M}; \citealt{2016ApJ...830...62M}; \citealt{2016ApJ...833..168M}; \citealt{2017ApJ...837..136D}). The surface brightness limit in this survey is $\sim 29.5 \textrm{ mag arcsec}^{-2}$ on scales of $\sim 10 \textrm{ arcsec}$.
For the purpose of comparing to star count surveys we adopt parameters of two surveys:  the Dark Energy Survey (DES; \citealt{2014SPIE.9149E..0VD} and references therein), performed with the Dark Energy Camera (DECam; \citealt{2010SPIE.7735E..0DF}; \citealt{2015AJ....150..150F}; \citealt{2012PhPro..37.1332D}) where a $g-$band limiting magnitude of $24.6 \textrm{ mag}$ is assumed (\citealt{2005astro.ph.10346T}) and the 
Hyper Supreme Cam Survey (\citealt{2017arXiv170208449A}) where a $g-$band limiting magnitude of $26.5 \textrm{ mag}$ is assumed.
For the two galaxy detection methods, integrated light and star counts, we calculate the limiting distance for detection as a function of stellar mass, i.e., the farthest distance a galaxy in a particular stellar mass is likely to be detected.

For the integrated light detections, the following steps were taken: for each stellar mass, $10^6$ galaxies were simulated, where for each galaxy a surface brightness and an effective radius (in kpc) were assigned such that they will be normally distributed with means and variances calculated in \ref{methods}, $\mu_\textrm{eff,V} \sim \mathcal{N}(\mu_{\mu_{\textrm{eff,V}}},\sigma_{\mu_{\textrm{eff,V}}}^2)$ and $R_\textrm{eff} \sim \mathcal{N}(\mu_{R_{\textrm{eff}}},\sigma_{R_{\textrm{eff}}}^2)$. The galaxies were then placed at random distances ($D=0-20\textrm{ Mpc}$) and their corresponding angular sizes in arcseconds were calculated. The integrated surface brightness is independent of distance for these distances. Galaxies with surface brightness lower than $29.5 \textrm{ mag arcsec}^{-2}$ and with sizes larger than $10 \textrm{ arcsec}$ were flagged as `detected'. Then, the limiting distance for detection was calculated for $20, 50 \textrm{ and } 80\%$ out of the total number of simulated galaxies. 

Detections using the star count method were determined in the following way: first, we calculate a MIST model isochrone for a single stellar population of age $10.0 \textrm{ Gyr}$ and metallicity of $\textrm{[Fe/H] = -2}$, and obtain synthetic photometry in the DECam bands. Given the galaxy stellar mass, the number of stars is obtained by integrating the IMF weights for those stars with magnitudes brighter than $g-$band magnitude limits of $24.6$ mag and $26.5$ mag. The limiting distance, $\textrm{D}_{\textrm{lim}}$ for the two star count surveys, is the distance for which a galaxy with stellar mass, $\textrm{M}_*$ has a minimal number of detectable stars with an apparent magnitude brighter than $24.6$ mag and $26.5$ mag, respectively. 
We assume that a minimum of $20 \pm 10$ stars is required for a significant detection. This is a conservative lower limit, based on inspection of color-magnitude diagrams in the discovery papers of Milky Way ultra faint dwarfs (\citealt{2014MNRAS.441.2124B}; \citealt{2016MNRAS.463..712T}).\footnote{The number of stars is usually not provided, but appears to be typically larger than 10 (Belokurov, private communication).} 
Also, we assume that all stars brighter than the survey limit can be used in the analysis. In practice, brighter limits are usually employed; as an example, \citet{2015ApJ...805..130K} used $r<23$ mag rather than $r<24.6$ mag due to uncertainties in the star/galaxy classification at fainter magnitudes in DES.

The resulting limiting distances as a function of stellar mass for integrated light and star counts detections are shown is Figure \ref{fig:comparison}. The blue shaded regions show detectable distances for $20-50-80\%$ of the galaxies in the simulated Dragonfly sample where the grey curves show the detectable distances for star count surveys with magnitude limits of $24.6$ mag and $26.5$ mag in $g-$band (Shaded grey regions show the effect of varying the minimum number of detected stars between 10 and 30).
An integrated light survey using the Dragonfly Telephoto Array is forecasted to reach greater limiting distances than using star count survey with the listed magnitude limits, for a large fraction of the simulated galaxies. 
The mass cutoff in the $80\%$ detection curve is due to Dragonfly's limiting surface brightness while the limiting distance for all of the Dragonfly curves is determined by the angular resolution. The shape of the star count surveys detections limit curve matches the general shape of the isochrones: we require 20 stars to be detected, and at low masses the galaxies do not have this many giants. As a result, the brightest stars are subgiants, and the limiting distance plummets for these faint stars. 
At the high mass end, star count surveys are restricted by the limiting magnitude of the instrument as the brightness of stars, and thus the number of detectable tracer stars, decreases with the square of the distance. In contrast, integrated light surface brightness is conserved with distance and thus integrated light surveys are mostly restricted by their limiting surface brightness and resolution, thus allow dwarfs to be detected beyond the Local Group.
The power of these complementary approaches can also be seen in Figure \ref{fig:cumulative}.
We show the cumulative number of expected dwarf galaxies in a Dragonfly integrated light survey and in a star count surveys assuming two magnitude limits. 
Similar to Section \ref{abundance}, we assume dark matter halo mass function from Tinker et al. (\citeyear{2010ApJ...724..878T}) and repeat the calculation for the two stellar mass-halo mass relation: B13 and RP17. As demonstrated before, the expected detection rate increases by an order of magnitude when adopting the B13 stellar mass-halo mass relation compared to RP17.
The two imaging techniques essentially cover complementary phase space where star count surveys are better at detecting the plentiful lower mass nearby galaxies, where as integrated light surveys help to increase the number of extended, low and high mass, galaxy candidates further away, depending mostly on their limiting surface brightness and spatial resolution.


\section{Summary and Discussion}\label{discussion}

Recent developments of integrated light techniques are sensitive enough to allow dwarfs to be detected beyond the Local Group (\citealt{2014arXiv1401.2719K}; \citealt{2014ApJ...787L..37M}) and to allow a statistical probe of the low mass dwarf population. 

In this paper we study the prospects of integrated light imaging in the context of constructing a complete census of dwarf galaxies in the Local Volume general field (between $3$ and $10$ Mpc), down to very low masses.

We present a model for calculating the predicted detection rates of dwarf galaxies using integrated light surveys, depending on their limiting effective surface brightness and spatial resolution. 
Two assumptions are made and should be noted:
\begin{enumerate}
  \item We partially base our model on properties that were measured for the dwarf galaxy population in the Local Group, which can ultimately be very different from the statistical properties of dwarf galaxies in the field. This assumption can be revisited when the first samples of very low mass field galaxies are available.
  \item We assume a one-to-one relation for the stellar mass-halo mass ($\textrm{M}_{*}-\textrm{M}_{h}$) relation where the scatter at the low mass end may be much larger than the 0.2 dex scatter measured at the high mass end. 
\end{enumerate}

The principle result of this paper is presented in Figure \ref{fig:abundances}. Assuming two $\textrm{M}_{*}-\textrm{M}_{h}$ relations, Behroozi et al. (\citeyear{2013ApJ...770...57B}) and the Rodrigues-Puebla et al. (\citeyear{2017arXiv170304542R}) ,we present the predicted abundances of field dwarfs in the Local Volume, in integrated light surveys over a range of values for the limiting surface brightness, $\mu_{\textrm{eff,lim}}$ and spatial resolution, $\theta$.
Assuming the B13 relation, low mass dwarf galaxies should be detected in large numbers, $\sim 0.3-0.4 \textrm{ deg}^{-2}$, when carrying out an integrated light survey with a limiting surface brightness larger than $\sim 29 \textrm{ mag arcsec}^{-2}$ and a spatial resolution better than $\sim 5 \textrm{ arcsec}$. The result decreases by an order of magnitude when we adopt the RP17 relation. 
This drastic change when adopting two different stellar mass-halo mass relations, illustrates the necessity of performing a systematic search of such objects in the field. Proving the existence or alternatively the lack of a large population of faint and ultra-faint galaxies will provide an important constraint on the $\textrm{M}_*-\textrm{M}_h$ relation at low masses. 

We compare our results to those that can be achieved using the `standard' star counts method in Figures \ref{fig:comparison} and \ref{fig:cumulative}. 
We demonstrate that integrated light imaging is complementary to the star counts method and has different strengths and weaknesses. While the star counts technique is dominated by the inverse square law, imaging the integrated light of extended galaxies takes advantage of the conservation of surface brightness in the Local Universe.
Motivated by the results of this study we are in the first stages of conducting the `Dragonfly Blank Wide Field Survey'. This is a deep photometric survey of a wide blank area to be carried out with the Dragonfly Telephoto Array. Its main goal is to detect a large set of galaxy candidates, predicted to exist in the field, and to study their properties. 
We hope to be able to shed light on theories of isolated galaxy formation which are currently fail to be constrained since so-far no low mass dwarfs were detected in the field.
In order to study these galaxy candidates further, follow-up deep high resolution observations will need to be taken.

\acknowledgments
\section*{Acknowledgement}

The authors thank Peter Berhoozi, Ferah Munshi and Aldo Rodriguez for providing the stellar mass-halo mass tables. We thank Marla Geha for valuable discussions and ideas.
Support from NSF grants AST-1312376 and AST-1613582 is gratefully acknowledged.

\bibliography{paper2_ref}

\begin{thebibliography}{}
\expandafter\ifx\csname natexlab\endcsname\relax\def\natexlab#1{#1}\fi
\providecommand{\url}[1]{\href{#1}{#1}}

\bibitem[{{Abraham} \& {van Dokkum}(2014)}]{2014PASP..126...55A}
{Abraham}, R.~G., \& {van Dokkum}, P.~G. 2014, \pasp, 126, 55

\bibitem[{{Aihara} {et~al.}(2017){Aihara}, {Armstrong}, {Bickerton}, {Bosch},
  {Coupon}, {Furusawa}, {Hayashi}, {Ikeda}, {Kamata}, {Karoji}, {Kawanomoto},
  {Koike}, {Komiyama}, {Lupton}, {Mineo}, {Miyatake}, {Miyazaki}, {Morokuma},
  {Obuchi}, {Oishi}, {Okura}, {Price}, {Takata}, {Tanaka}, {Tanaka}, {Tanaka},
  {Uchida}, {Uraguchi}, {Utsumi}, {Wang}, {Yamada}, {Yamanoi}, {Yasuda},
  {Arimoto}, {Chiba}, {Finet}, {Fujimori}, {Fujimoto}, {Furusawa}, {Goto},
  {Goulding}, {Gunn}, {Harikane}, {Hattori}, {Hayashi}, {Helminiak}, {Higuchi},
  {Hikage}, {Ho}, {Hsieh}, {Huang}, {Huang}, {Imanishi}, {Iwata}, {Jaelani},
  {Jian}, {Kashikawa}, {Katayama}, {Kojima}, {Konno}, {Koshida}, {Kusakabe},
  {Leauthaud}, {Lee}, {Lin}, {Lin}, {Mandelbaum}, {Matsuoka}, {Medezinski},
  {Miyama}, {Momose}, {More}, {More}, {Mukae}, {Murata}, {Murayama}, {Nagao},
  {Nakata}, {Niikura}, {Nishizawa}, {Oguri}, {Okabe}, {Ono}, {Onodera},
  {Onoue}, {Ouchi}, {Pyo}, {Shibuya}, {Shimasaku}, {Simet}, {Speagle},
  {Spergel}, {Strauss}, {Sugahara}, {Sugiyama}, {Suto}, {Suzuki}, {Tait},
  {Takada}, {Terai}, {Toba}, {Turner}, {Uchiyama}, {Umetsu}, {Urata}, {Usuda},
  {Yeh}, \& {Yuma}}]{2017arXiv170208449A}
{Aihara}, H., {Armstrong}, R., {Bickerton}, S., {et~al.} 2017, ArXiv e-prints,
  arXiv:1702.08449

\bibitem[{{Bechtol} {et~al.}(2015){Bechtol}, {Drlica-Wagner}, {Balbinot},
  {Pieres}, {Simon}, {Yanny}, {Santiago}, {Wechsler}, {Frieman}, {Walker},
  {Williams}, {Rozo}, {Rykoff}, {Queiroz}, {Luque}, {Benoit-L{\'e}vy},
  {Tucker}, {Sevilla}, {Gruendl}, {da Costa}, {Fausti Neto}, {Maia}, {Abbott},
  {Allam}, {Armstrong}, {Bauer}, {Bernstein}, {Bernstein}, {Bertin}, {Brooks},
  {Buckley-Geer}, {Burke}, {Carnero Rosell}, {Castander}, {Covarrubias},
  {D'Andrea}, {DePoy}, {Desai}, {Diehl}, {Eifler}, {Estrada}, {Evrard},
  {Fernandez}, {Finley}, {Flaugher}, {Gaztanaga}, {Gerdes}, {Girardi},
  {Gladders}, {Gruen}, {Gutierrez}, {Hao}, {Honscheid}, {Jain}, {James},
  {Kent}, {Kron}, {Kuehn}, {Kuropatkin}, {Lahav}, {Li}, {Lin}, {Makler},
  {March}, {Marshall}, {Martini}, {Merritt}, {Miller}, {Miquel}, {Mohr},
  {Neilsen}, {Nichol}, {Nord}, {Ogando}, {Peoples}, {Petravick}, {Plazas},
  {Romer}, {Roodman}, {Sako}, {Sanchez}, {Scarpine}, {Schubnell}, {Smith},
  {Soares-Santos}, {Sobreira}, {Suchyta}, {Swanson}, {Tarle}, {Thaler},
  {Thomas}, {Wester}, {Zuntz}, \& {DES Collaboration}}]{2015ApJ...807...50B}
{Bechtol}, K., {Drlica-Wagner}, A., {Balbinot}, E., {et~al.} 2015, \apj, 807,
  50

\bibitem[{{Behroozi} {et~al.}(2013){Behroozi}, {Wechsler}, \&
  {Conroy}}]{2013ApJ...770...57B}
{Behroozi}, P.~S., {Wechsler}, R.~H., \& {Conroy}, C. 2013, \apj, 770, 57

\bibitem[{{Belokurov} {et~al.}(2014){Belokurov}, {Irwin}, {Koposov}, {Evans},
  {Gonzalez-Solares}, {Metcalfe}, \& {Shanks}}]{2014MNRAS.441.2124B}
{Belokurov}, V., {Irwin}, M.~J., {Koposov}, S.~E., {et~al.} 2014, \mnras, 441,
  2124

\bibitem[{{Belokurov} {et~al.}(2010){Belokurov}, {Walker}, {Evans}, {Gilmore},
  {Irwin}, {Just}, {Koposov}, {Mateo}, {Olszewski}, {Watkins}, \&
  {Wyrzykowski}}]{2010ApJ...712L.103B}
{Belokurov}, V., {Walker}, M.~G., {Evans}, N.~W., {et~al.} 2010, \apjl, 712,
  L103

\bibitem[{{Brook} {et~al.}(2014){Brook}, {Di Cintio}, {Knebe}, {Gottl{\"o}ber},
  {Hoffman}, {Yepes}, \& {Garrison-Kimmel}}]{2014ApJ...784L..14B}
{Brook}, C.~B., {Di Cintio}, A., {Knebe}, A., {et~al.} 2014, \apjl, 784, L14

\bibitem[{{Choi} {et~al.}(2016){Choi}, {Dotter}, {Conroy}, {Cantiello},
  {Paxton}, \& {Johnson}}]{2016ApJ...823..102C}
{Choi}, J., {Dotter}, A., {Conroy}, C., {et~al.} 2016, \apj, 823, 102

\bibitem[{{Conroy} {et~al.}(2009){Conroy}, {Gunn}, \&
  {White}}]{2009ApJ...699..486C}
{Conroy}, C., {Gunn}, J.~E., \& {White}, M. 2009, \apj, 699, 486

\bibitem[{{Conroy} \& {van Dokkum}(2012)}]{2012ApJ...747...69C}
{Conroy}, C., \& {van Dokkum}, P. 2012, \apj, 747, 69

\bibitem[{{Conroy} {et~al.}(2006){Conroy}, {Wechsler}, \&
  {Kravtsov}}]{2006ApJ...647..201C}
{Conroy}, C., {Wechsler}, R.~H., \& {Kravtsov}, A.~V. 2006, \apj, 647, 201

\bibitem[{{Danieli} {et~al.}(2017){Danieli}, {van Dokkum}, {Merritt},
  {Abraham}, {Zhang}, {Karachentsev}, \& {Makarova}}]{2017ApJ...837..136D}
{Danieli}, S., {van Dokkum}, P., {Merritt}, A., {et~al.} 2017, \apj, 837, 136

\bibitem[{{Diehl} {et~al.}(2014){Diehl}, {Abbott}, {Annis}, {Armstrong},
  {Baruah}, {Bermeo}, {Bernstein}, {Beynon}, {Bruderer}, {Buckley-Geer},
  {Campbell}, {Capozzi}, {Carter}, {Casas}, {Clerkin}, {Covarrubias}, {Cuhna},
  {D'Andrea}, {da Costa}, {Das}, {DePoy}, {Dietrich}, {Drlica-Wagner},
  {Elliott}, {Eifler}, {Estrada}, {Etherington}, {Flaugher}, {Frieman}, {Fausti
  Neto}, {Gelman}, {Gerdes}, {Gruen}, {Gruendl}, {Hao}, {Head}, {Helsby},
  {Hoffman}, {Honscheid}, {James}, {Johnson}, {Kacprzac}, {Katsaros},
  {Kennedy}, {Kent}, {Kessler}, {Kim}, {Krause}, {Kron}, {Kuhlmann}, {Kunder},
  {Li}, {Lin}, {Maccrann}, {March}, {Marshall}, {Neilsen}, {Nugent}, {Martini},
  {Melchior}, {Menanteau}, {Nichol}, {Nord}, {Ogando}, {Old}, {Papadopoulos},
  {Patton}, {Petravick}, {Plazas}, {Poulton}, {Pujol}, {Reil}, {Rigby},
  {Romer}, {Roodman}, {Rooney}, {Sanchez Alvaro}, {Serrano}, {Sheldon},
  {Smith}, {Smith}, {Soares-Santos}, {Soumagnac}, {Spinka}, {Suchyta},
  {Tucker}, {Walker}, {Wester}, {Wiesner}, {Wilcox}, {Williams}, {Yanny}, \&
  {Zhang}}]{2014SPIE.9149E..0VD}
{Diehl}, H.~T., {Abbott}, T.~M.~C., {Annis}, J., {et~al.} 2014, in \procspie,
  Vol. 9149, Observatory Operations: Strategies, Processes, and Systems V,
  91490V

\bibitem[{{Diehl} \& {Dark Energy Survey
  Collaboration}(2012)}]{2012PhPro..37.1332D}
{Diehl}, T., \& {Dark Energy Survey Collaboration}. 2012, Physics Procedia, 37,
  1332

\bibitem[{{Dotter}(2016)}]{2016ApJS..222....8D}
{Dotter}, A. 2016, \apjs, 222, 8

\bibitem[{{Drlica-Wagner} {et~al.}(2015){Drlica-Wagner}, {Bechtol}, {Rykoff},
  {Luque}, {Queiroz}, {Mao}, {Wechsler}, {Simon}, {Santiago}, {Yanny},
  {Balbinot}, {Dodelson}, {Fausti Neto}, {James}, {Li}, {Maia}, {Marshall},
  {Pieres}, {Stringer}, {Walker}, {Abbott}, {Abdalla}, {Allam},
  {Benoit-L{\'e}vy}, {Bernstein}, {Bertin}, {Brooks}, {Buckley-Geer}, {Burke},
  {Carnero Rosell}, {Carrasco Kind}, {Carretero}, {Crocce}, {da Costa},
  {Desai}, {Diehl}, {Dietrich}, {Doel}, {Eifler}, {Evrard}, {Finley},
  {Flaugher}, {Fosalba}, {Frieman}, {Gaztanaga}, {Gerdes}, {Gruen}, {Gruendl},
  {Gutierrez}, {Honscheid}, {Kuehn}, {Kuropatkin}, {Lahav}, {Martini},
  {Miquel}, {Nord}, {Ogando}, {Plazas}, {Reil}, {Roodman}, {Sako}, {Sanchez},
  {Scarpine}, {Schubnell}, {Sevilla-Noarbe}, {Smith}, {Soares-Santos},
  {Sobreira}, {Suchyta}, {Swanson}, {Tarle}, {Tucker}, {Vikram}, {Wester},
  {Zhang}, {Zuntz}, \& {DES Collaboration}}]{2015ApJ...813..109D}
{Drlica-Wagner}, A., {Bechtol}, K., {Rykoff}, E.~S., {et~al.} 2015, \apj, 813,
  109

\bibitem[{{Ferrarese} {et~al.}(2016){Ferrarese}, {C{\^o}t{\'e}},
  {S{\'a}nchez-Janssen}, {Roediger}, {McConnachie}, {Durrell}, {MacArthur},
  {Blakeslee}, {Duc}, {Boissier}, {Boselli}, {Courteau}, {Cuillandre},
  {Emsellem}, {Gwyn}, {Guhathakurta}, {Jord{\'a}n}, {Lan{\c c}on}, {Liu},
  {Mei}, {Mihos}, {Navarro}, {Peng}, {Puzia}, {Taylor}, {Toloba}, \&
  {Zhang}}]{2016ApJ...824...10F}
{Ferrarese}, L., {C{\^o}t{\'e}}, P., {S{\'a}nchez-Janssen}, R., {et~al.} 2016,
  \apj, 824, 10

\bibitem[{{Flaugher} {et~al.}(2015){Flaugher}, {Diehl}, {Honscheid}, {Abbott},
  {Alvarez}, {Angstadt}, {Annis}, {Antonik}, {Ballester}, {Beaufore},
  {Bernstein}, {Bernstein}, {Bigelow}, {Bonati}, {Boprie}, {Brooks},
  {Buckley-Geer}, {Campa}, {Cardiel-Sas}, {Castander}, {Castilla}, {Cease},
  {Cela-Ruiz}, {Chappa}, {Chi}, {Cooper}, {da Costa}, {Dede}, {Derylo},
  {DePoy}, {de Vicente}, {Doel}, {Drlica-Wagner}, {Eiting}, {Elliott}, {Emes},
  {Estrada}, {Fausti Neto}, {Finley}, {Flores}, {Frieman}, {Gerdes},
  {Gladders}, {Gregory}, {Gutierrez}, {Hao}, {Holland}, {Holm}, {Huffman},
  {Jackson}, {James}, {Jonas}, {Karcher}, {Karliner}, {Kent}, {Kessler},
  {Kozlovsky}, {Kron}, {Kubik}, {Kuehn}, {Kuhlmann}, {Kuk}, {Lahav}, {Lathrop},
  {Lee}, {Levi}, {Lewis}, {Li}, {Mandrichenko}, {Marshall}, {Martinez},
  {Merritt}, {Miquel}, {Mu{\~n}oz}, {Neilsen}, {Nichol}, {Nord}, {Ogando},
  {Olsen}, {Palaio}, {Patton}, {Peoples}, {Plazas}, {Rauch}, {Reil}, {Rheault},
  {Roe}, {Rogers}, {Roodman}, {Sanchez}, {Scarpine}, {Schindler}, {Schmidt},
  {Schmitt}, {Schubnell}, {Schultz}, {Schurter}, {Scott}, {Serrano}, {Shaw},
  {Smith}, {Soares-Santos}, {Stefanik}, {Stuermer}, {Suchyta}, {Sypniewski},
  {Tarle}, {Thaler}, {Tighe}, {Tran}, {Tucker}, {Walker}, {Wang}, {Watson},
  {Weaverdyck}, {Wester}, {Woods}, {Yanny}, \& {DES
  Collaboration}}]{2015AJ....150..150F}
{Flaugher}, B., {Diehl}, H.~T., {Honscheid}, K., {et~al.} 2015, \aj, 150, 150

\bibitem[{{Flaugher} {et~al.}(2010){Flaugher}, {Abbott}, {Annis}, {Antonik},
  {Bailey}, {Ballester}, {Bernstein}, {Bernstein}, {Bonati}, {Bremer},
  {Briones}, {Brooks}, {Buckley-Geer}, {Campa}, {Cardiel-Sas}, {Castander},
  {Castilla}, {Cease}, {Chappa}, {Chi}, {da Costa}, {DePoy}, {Derylo}, {De
  Vicente}, {Diehl}, {Doel}, {Estrada}, {Eiting}, {Elliott}, {Finley},
  {Frieman}, {Gaztanaga}, {Gerdes}, {Gladders}, {Guarino}, {Gutierrez},
  {Grudzinski}, {Hanlon}, {Hao}, {Holland}, {Honscheid}, {Huffman}, {Jackson},
  {Karliner}, {Kau}, {Kent}, {Krempetz}, {Krider}, {Kozlovsky}, {Kubik},
  {Kuehn}, {Kuhlmann}, {Kuk}, {Lahav}, {Lewis}, {Lin}, {Lorenzon}, {Marshall},
  {Mart{\'{\i}}nez}, {McKay}, {Merritt}, {Meyer}, {Miquel}, {Morgan}, {Moore},
  {Moore}, {Nord}, {Ogando}, {Olsen}, {Peoples}, {Plazas}, {Roe}, {Roodman},
  {Rossetto}, {Sanchez}, {Scarpine}, {Schalk}, {Schindler}, {Schmidt},
  {Schmitt}, {Schubnell}, {Schultz}, {Selen}, {Serrano}, {Shaw}, {Simaitis},
  {Slaughter}, {Smith}, {Spinka}, {Stefanik}, {Stuermer}, {Sypniewski},
  {Talaga}, {Tarle}, {Thaler}, {Tucker}, {Walker}, {Weaverdyck}, {Wester},
  {Woods}, {Worswick}, \& {Zhao}}]{2010SPIE.7735E..0DF}
{Flaugher}, B.~L., {Abbott}, T.~M.~C., {Annis}, J., {et~al.} 2010, in
  \procspie, Vol. 7735, Ground-based and Airborne Instrumentation for Astronomy
  III, 77350D

\bibitem[{{Frenk} {et~al.}(1988){Frenk}, {White}, {Davis}, \&
  {Efstathiou}}]{1988ApJ...327..507F}
{Frenk}, C.~S., {White}, S.~D.~M., {Davis}, M., \& {Efstathiou}, G. 1988, \apj,
  327, 507

\bibitem[{{Garrison-Kimmel} {et~al.}(2014){Garrison-Kimmel}, {Boylan-Kolchin},
  {Bullock}, \& {Lee}}]{2014MNRAS.438.2578G}
{Garrison-Kimmel}, S., {Boylan-Kolchin}, M., {Bullock}, J.~S., \& {Lee}, K.
  2014, \mnras, 438, 2578

\bibitem[{{Garrison-Kimmel} {et~al.}(2017){Garrison-Kimmel}, {Bullock},
  {Boylan-Kolchin}, \& {Bardwell}}]{2017MNRAS.464.3108G}
{Garrison-Kimmel}, S., {Bullock}, J.~S., {Boylan-Kolchin}, M., \& {Bardwell},
  E. 2017, \mnras, 464, 3108

\bibitem[{{Geha} {et~al.}(2012){Geha}, {Blanton}, {Yan}, \&
  {Tinker}}]{2012ApJ...757...85G}
{Geha}, M., {Blanton}, M.~R., {Yan}, R., \& {Tinker}, J.~L. 2012, \apj, 757, 85

\bibitem[{{Geha} {et~al.}(2009){Geha}, {Willman}, {Simon}, {Strigari}, {Kirby},
  {Law}, \& {Strader}}]{2009ApJ...692.1464G}
{Geha}, M., {Willman}, B., {Simon}, J.~D., {et~al.} 2009, \apj, 692, 1464

\bibitem[{{Geha} {et~al.}(2017){Geha}, {Wechsler}, {Mao}, {Tollerud}, {Weiner},
  {Bernstein}, {Hoyle}, {Marchi}, {Marshall}, {Mu{\~n}oz}, \&
  {Lu}}]{2017ApJ...847....4G}
{Geha}, M., {Wechsler}, R.~H., {Mao}, Y.-Y., {et~al.} 2017, \apj, 847, 4

\bibitem[{{Greco} {et~al.}(2017){Greco}, {Greene}, {Strauss}, {MacArthur},
  {Flowers}, {Goulding}, {Huang}, {Kim}, {Komiyama}, {Leauthaud}, {Leisman},
  {Lupton}, {Sif{\'o}n}, \& {Wang}}]{2017arXiv170904474G}
{Greco}, J.~P., {Greene}, J.~E., {Strauss}, M.~A., {et~al.} 2017, ArXiv
  e-prints, arXiv:1709.04474

\bibitem[{{Guo} {et~al.}(2010){Guo}, {White}, {Li}, \&
  {Boylan-Kolchin}}]{2010MNRAS.404.1111G}
{Guo}, Q., {White}, S., {Li}, C., \& {Boylan-Kolchin}, M. 2010, \mnras, 404,
  1111

\bibitem[{{Henkel} {et~al.}(2017){Henkel}, {Javanmardi},
  {Mart{\'{\i}}nez-Delgado}, {Kroupa}, \& {Teuwen}}]{2017A&A...603A..18H}
{Henkel}, C., {Javanmardi}, B., {Mart{\'{\i}}nez-Delgado}, D., {Kroupa}, P., \&
  {Teuwen}, K. 2017, \aap, 603, A18

\bibitem[{{Homma} {et~al.}(2016){Homma}, {Chiba}, {Okamoto}, {Komiyama},
  {Tanaka}, {Tanaka}, {Ishigaki}, {Akiyama}, {Arimoto}, {Garmilla}, {Lupton},
  {Strauss}, {Furusawa}, {Miyazaki}, {Murayama}, {Nishizawa}, {Takada},
  {Usuda}, \& {Wang}}]{2016ApJ...832...21H}
{Homma}, D., {Chiba}, M., {Okamoto}, S., {et~al.} 2016, \apj, 832, 21

\bibitem[{{Homma} {et~al.}(2017){Homma}, {Chiba}, {Okamoto}, {Komiyama},
  {Tanaka}, {Tanaka}, {Ishigaki}, {Hayashi}, {Arimoto}, {Garmilla}, {Lupton},
  {Strauss}, {Miyazaki}, {Wang}, \& {Murayama}}]{2017arXiv170405977H}
---. 2017, ArXiv e-prints, arXiv:1704.05977

\bibitem[{{Ibata} {et~al.}(2007){Ibata}, {Martin}, {Irwin}, {Chapman},
  {Ferguson}, {Lewis}, \& {McConnachie}}]{2007ApJ...671.1591I}
{Ibata}, R., {Martin}, N.~F., {Irwin}, M., {et~al.} 2007, \apj, 671, 1591

\bibitem[{{Irwin}(1994)}]{1994ESOC...49...27I}
{Irwin}, M.~J. 1994, in European Southern Observatory Conference and Workshop
  Proceedings, Vol.~49, European Southern Observatory Conference and Workshop
  Proceedings, ed. G.~{Meylan} \& P.~{Prugniel}, 27

\bibitem[{{Javanmardi} {et~al.}(2016){Javanmardi}, {Martinez-Delgado},
  {Kroupa}, {Henkel}, {Crawford}, {Teuwen}, {Gabany}, {Hanson}, {Chonis}, \&
  {Neyer}}]{2016A&A...588A..89J}
{Javanmardi}, B., {Martinez-Delgado}, D., {Kroupa}, P., {et~al.} 2016, \aap,
  588, A89

\bibitem[{{Jethwa} {et~al.}(2018){Jethwa}, {Erkal}, \&
  {Belokurov}}]{2018MNRAS.473.2060J}
{Jethwa}, P., {Erkal}, D., \& {Belokurov}, V. 2018, \mnras, 473, 2060

\bibitem[{{Karachentsev} {et~al.}(2014){Karachentsev}, {Bautzmann}, {Neyer},
  {Polzl}, {Riepe}, {Zilch}, \& {Mattern}}]{2014arXiv1401.2719K}
{Karachentsev}, I.~D., {Bautzmann}, D., {Neyer}, F., {et~al.} 2014, ArXiv
  e-prints, arXiv:1401.2719

\bibitem[{{Karachentsev} {et~al.}(2013){Karachentsev}, {Makarov}, \&
  {Kaisina}}]{2013AJ....145..101K}
{Karachentsev}, I.~D., {Makarov}, D.~I., \& {Kaisina}, E.~I. 2013, \aj, 145,
  101

\bibitem[{{Karachentsev} {et~al.}(2015){Karachentsev}, {Riepe}, {Zilch},
  {Blauensteiner}, {Elvov}, {Hochleitner}, {Hubl}, {Kerschhuber},
  {K{\"u}ppers}, {Neyer}, {P{\"o}lzl}, {Remmel}, {Schneider}, {Sparenberg},
  {Trulson}, {Willems}, \& {Ziegler}}]{2015AstBu..70..379K}
{Karachentsev}, I.~D., {Riepe}, P., {Zilch}, T., {et~al.} 2015, Astrophysical
  Bulletin, 70, 379

\bibitem[{{Karukes} \& {Salucci}(2017)}]{2017MNRAS.465.4703K}
{Karukes}, E.~V., \& {Salucci}, P. 2017, \mnras, 465, 4703

\bibitem[{{Klypin} {et~al.}(2015){Klypin}, {Karachentsev}, {Makarov}, \&
  {Nasonova}}]{2015MNRAS.454.1798K}
{Klypin}, A., {Karachentsev}, I., {Makarov}, D., \& {Nasonova}, O. 2015,
  \mnras, 454, 1798

\bibitem[{{Komatsu} {et~al.}(2011){Komatsu}, {Smith}, {Dunkley}, {Bennett},
  {Gold}, {Hinshaw}, {Jarosik}, {Larson}, {Nolta}, {Page}, {Spergel},
  {Halpern}, {Hill}, {Kogut}, {Limon}, {Meyer}, {Odegard}, {Tucker}, {Weiland},
  {Wollack}, \& {Wright}}]{2011ApJS..192...18K}
{Komatsu}, E., {Smith}, K.~M., {Dunkley}, J., {et~al.} 2011, \apjs, 192, 18

\bibitem[{{Koposov} {et~al.}(2008){Koposov}, {Belokurov}, {Evans}, {Hewett},
  {Irwin}, {Gilmore}, {Zucker}, {Rix}, {Fellhauer}, {Bell}, \&
  {Glushkova}}]{2008ApJ...686..279K}
{Koposov}, S., {Belokurov}, V., {Evans}, N.~W., {et~al.} 2008, \apj, 686, 279

\bibitem[{{Koposov} {et~al.}(2015){Koposov}, {Belokurov}, {Torrealba}, \&
  {Evans}}]{2015ApJ...805..130K}
{Koposov}, S.~E., {Belokurov}, V., {Torrealba}, G., \& {Evans}, N.~W. 2015,
  \apj, 805, 130

\bibitem[{{Kravtsov} {et~al.}(2004){Kravtsov}, {Berlind}, {Wechsler}, {Klypin},
  {Gottl{\"o}ber}, {Allgood}, \& {Primack}}]{2004ApJ...609...35K}
{Kravtsov}, A.~V., {Berlind}, A.~A., {Wechsler}, R.~H., {et~al.} 2004, \apj,
  609, 35

\bibitem[{{Martin} {et~al.}(2013){Martin}, {Slater}, {Schlafly}, {Morganson},
  {Rix}, {Bell}, {Laevens}, {Bernard}, {Ferguson}, {Finkbeiner}, {Burgett},
  {Chambers}, {Hodapp}, {Kaiser}, {Kudritzki}, {Magnier}, {Morgan}, {Price},
  {Tonry}, \& {Wainscoat}}]{2013ApJ...772...15M}
{Martin}, N.~F., {Slater}, C.~T., {Schlafly}, E.~F., {et~al.} 2013, \apj, 772,
  15

\bibitem[{{McConnachie}(2012)}]{2012AJ....144....4M}
{McConnachie}, A.~W. 2012, \aj, 144, 4

\bibitem[{{Merritt} {et~al.}(2014){Merritt}, {van Dokkum}, \&
  {Abraham}}]{2014ApJ...787L..37M}
{Merritt}, A., {van Dokkum}, P., \& {Abraham}, R. 2014, \apjl, 787, L37

\bibitem[{{Merritt} {et~al.}(2016{\natexlab{a}}){Merritt}, {van Dokkum},
  {Abraham}, \& {Zhang}}]{2016ApJ...830...62M}
{Merritt}, A., {van Dokkum}, P., {Abraham}, R., \& {Zhang}, J.
  2016{\natexlab{a}}, \apj, 830, 62

\bibitem[{{Merritt} {et~al.}(2016{\natexlab{b}}){Merritt}, {van Dokkum},
  {Danieli}, {Abraham}, {Zhang}, {Karachentsev}, \&
  {Makarova}}]{2016ApJ...833..168M}
{Merritt}, A., {van Dokkum}, P., {Danieli}, S., {et~al.} 2016{\natexlab{b}},
  \apj, 833, 168

\bibitem[{{Moster} {et~al.}(2013){Moster}, {Naab}, \&
  {White}}]{2013MNRAS.428.3121M}
{Moster}, B.~P., {Naab}, T., \& {White}, S.~D.~M. 2013, \mnras, 428, 3121

\bibitem[{{Moster} {et~al.}(2017){Moster}, {Naab}, \&
  {White}}]{2017arXiv170505373M}
---. 2017, ArXiv e-prints, arXiv:1705.05373

\bibitem[{{M{\"u}ller} {et~al.}(2017){M{\"u}ller}, {Scalera}, {Binggeli}, \&
  {Jerjen}}]{2017arXiv170103681M}
{M{\"u}ller}, O., {Scalera}, R., {Binggeli}, B., \& {Jerjen}, H. 2017, ArXiv
  e-prints, arXiv:1701.03681

\bibitem[{{Munshi} {et~al.}(2017){Munshi}, {Brooks}, {Applebaum}, {Weisz},
  {Governato}, \& {Quinn}}]{2017arXiv170506286M}
{Munshi}, F., {Brooks}, A.~M., {Applebaum}, E., {et~al.} 2017, ArXiv e-prints,
  arXiv:1705.06286

\bibitem[{{Murray} {et~al.}(2013){Murray}, {Power}, \&
  {Robotham}}]{2013A&C.....3...23M}
{Murray}, S.~G., {Power}, C., \& {Robotham}, A.~S.~G. 2013, Astronomy and
  Computing, 3, 23

\bibitem[{{Read} {et~al.}(2017){Read}, {Iorio}, {Agertz}, \&
  {Fraternali}}]{2017MNRAS.467.2019R}
{Read}, J.~I., {Iorio}, G., {Agertz}, O., \& {Fraternali}, F. 2017, \mnras,
  467, 2019

\bibitem[{{Richardson} {et~al.}(2011){Richardson}, {Irwin}, {McConnachie},
  {Martin}, {Dotter}, {Ferguson}, {Ibata}, {Chapman}, {Lewis}, {Tanvir}, \&
  {Rich}}]{2011ApJ...732...76R}
{Richardson}, J.~C., {Irwin}, M.~J., {McConnachie}, A.~W., {et~al.} 2011, \apj,
  732, 76

\bibitem[{{Rodriguez-Puebla} {et~al.}(2017){Rodriguez-Puebla}, {Primack},
  {Avila-Reese}, \& {Faber}}]{2017arXiv170304542R}
{Rodriguez-Puebla}, A., {Primack}, J.~R., {Avila-Reese}, V., \& {Faber}, S.~M.
  2017, ArXiv e-prints, arXiv:1703.04542

\bibitem[{{Romanowsky} {et~al.}(2016){Romanowsky}, {Mart{\'{\i}}nez-Delgado},
  {Martin}, {Morales}, {Jennings}, {GaBany}, {Brodie}, {Grebel}, {Schedler}, \&
  {Sidonio}}]{2016MNRAS.457L.103R}
{Romanowsky}, A.~J., {Mart{\'{\i}}nez-Delgado}, D., {Martin}, N.~F., {et~al.}
  2016, \mnras, 457, L103

\bibitem[{{Salpeter}(1955)}]{1955ApJ...121..161S}
{Salpeter}, E.~E. 1955, \apj, 121, 161

\bibitem[{{Sawala} {et~al.}(2015){Sawala}, {Frenk}, {Fattahi}, {Navarro},
  {Bower}, {Crain}, {Dalla Vecchia}, {Furlong}, {Jenkins}, {McCarthy}, {Qu},
  {Schaller}, {Schaye}, \& {Theuns}}]{2015MNRAS.448.2941S}
{Sawala}, T., {Frenk}, C.~S., {Fattahi}, A., {et~al.} 2015, \mnras, 448, 2941

\bibitem[{{Simon} {et~al.}(2017){Simon}, {Li}, {Drlica-Wagner}, {Bechtol},
  {Marshall}, {James}, {Wang}, {Strigari}, {Balbinot}, {Kuehn}, {Walker},
  {Abbott}, {Allam}, {Annis}, {Benoit-L{\'e}vy}, {Brooks}, {Buckley-Geer},
  {Burke}, {Carnero Rosell}, {Carrasco Kind}, {Carretero}, {Cunha}, {D'Andrea},
  {da Costa}, {DePoy}, {Desai}, {Doel}, {Fernandez}, {Flaugher}, {Frieman},
  {Garc{\'{\i}}a-Bellido}, {Gaztanaga}, {Goldstein}, {Gruen}, {Gutierrez},
  {Kuropatkin}, {Maia}, {Martini}, {Menanteau}, {Miller}, {Miquel}, {Neilsen},
  {Nord}, {Ogando}, {Plazas}, {Romer}, {Rykoff}, {Sanchez}, {Santiago},
  {Scarpine}, {Schubnell}, {Sevilla-Noarbe}, {Smith}, {Sobreira}, {Suchyta},
  {Swanson}, {Tarle}, {Whiteway}, {Yanny}, \& {DES
  Collaboration}}]{2017ApJ...838...11S}
{Simon}, J.~D., {Li}, T.~S., {Drlica-Wagner}, A., {et~al.} 2017, \apj, 838, 11

\bibitem[{{The Dark Energy Survey Collaboration}(2005)}]{2005astro.ph.10346T}
{The Dark Energy Survey Collaboration}. 2005, ArXiv Astrophysics e-prints,
  astro-ph/0510346

\bibitem[{{Tinker} {et~al.}(2010){Tinker}, {Robertson}, {Kravtsov}, {Klypin},
  {Warren}, {Yepes}, \& {Gottl{\"o}ber}}]{2010ApJ...724..878T}
{Tinker}, J.~L., {Robertson}, B.~E., {Kravtsov}, A.~V., {et~al.} 2010, \apj,
  724, 878

\bibitem[{{Torrealba} {et~al.}(2016{\natexlab{a}}){Torrealba}, {Koposov},
  {Belokurov}, \& {Irwin}}]{2016MNRAS.459.2370T}
{Torrealba}, G., {Koposov}, S.~E., {Belokurov}, V., \& {Irwin}, M.
  2016{\natexlab{a}}, \mnras, 459, 2370

\bibitem[{{Torrealba} {et~al.}(2016{\natexlab{b}}){Torrealba}, {Koposov},
  {Belokurov}, {Irwin}, {Collins}, {Spencer}, {Ibata}, {Mateo}, {Bonaca}, \&
  {Jethwa}}]{2016MNRAS.463..712T}
{Torrealba}, G., {Koposov}, S.~E., {Belokurov}, V., {et~al.}
  2016{\natexlab{b}}, \mnras, 463, 712

\bibitem[{{Vale} \& {Ostriker}(2006)}]{2006MNRAS.371.1173V}
{Vale}, A., \& {Ostriker}, J.~P. 2006, \mnras, 371, 1173

\bibitem[{{Walsh} {et~al.}(2009){Walsh}, {Willman}, \&
  {Jerjen}}]{2009AJ....137..450W}
{Walsh}, S.~M., {Willman}, B., \& {Jerjen}, H. 2009, \aj, 137, 450

\bibitem[{{Yang} {et~al.}(2003){Yang}, {Mo}, \& {van den
  Bosch}}]{2003MNRAS.339.1057Y}
{Yang}, X., {Mo}, H.~J., \& {van den Bosch}, F.~C. 2003, \mnras, 339, 1057

\end{thebibliography}

\appendix 
\section{Dwarf galaxies with younger stellar populations}

\begin{figure*}[t]
\centering
  \includegraphics[width=\textwidth]{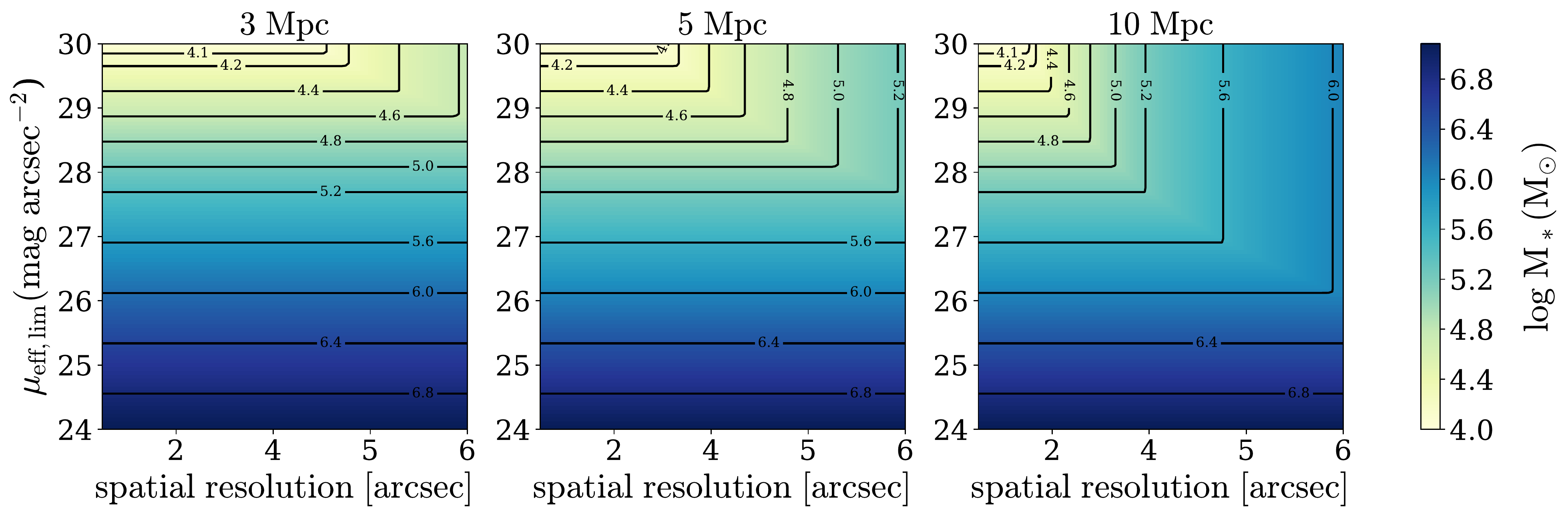}
  \caption{The minimal detectable stellar mass in integrated light surveys, assuming a limiting effective surface brightness in the $V-$band and a spatial resolution. Black contour lines indicate constant minimal stellar mass for different values of limiting effective surface brightness and spatial resolution. A $V-$band mass-to-light ratio of $M/L_{V} = \textrm{0.3}$ has assumed, appropriate for, e.g., a 1 Gyr old, Z/H=-1 stellar population.}
  \label{fig:min_stellar_mass_young}
\end{figure*}

\begin{figure*}
\centering
\subfigure
{
   \includegraphics[width=\textwidth]{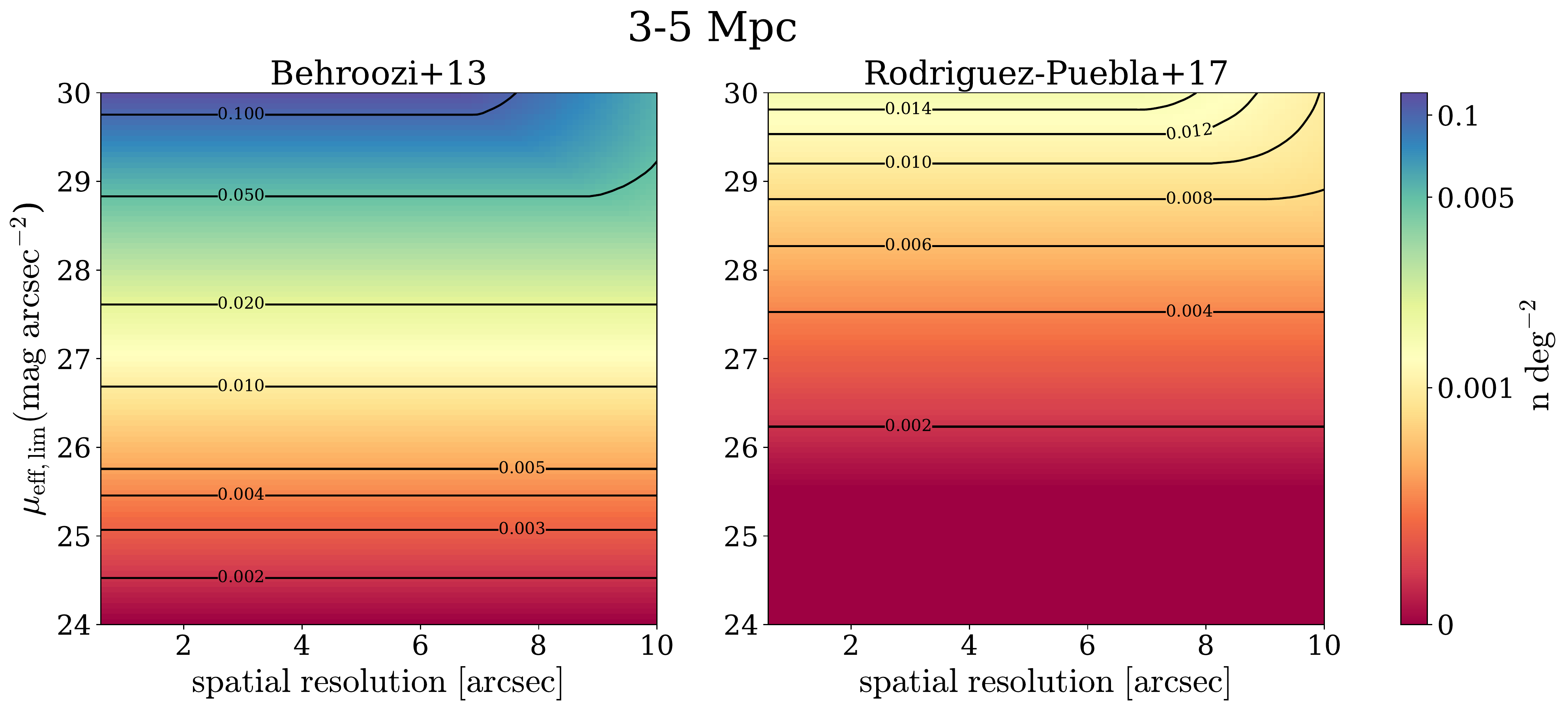}
   \label{fig:Cum1} 
}

\subfigure
{
   \includegraphics[width=\textwidth]{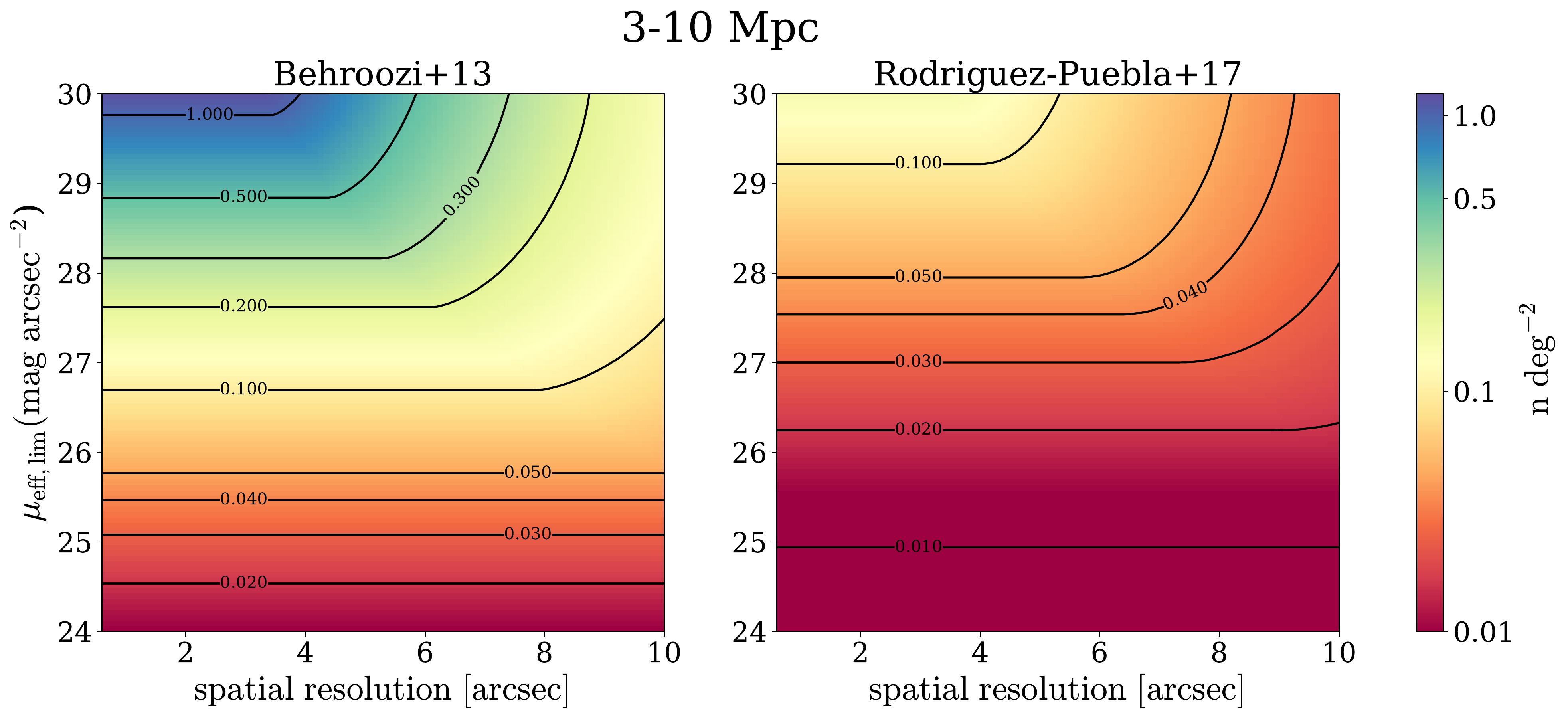}
   \label{fig:Cum2}
}
\caption{The cumulative number of predicted field galaxies per square degree to be detected in the Local Volume, between $3$ and $10$ Mpc (upper panel) and between $3$ and $5$ Mpc (lower panel) using an integrated light imaging, assuming a limiting effective surface brightness in $V$-band, $\mu_\textrm{eff,lim}$, and a spatial resolution, $\theta$. The left and right panels were calculated assuming the Behroozi et al. (\citeyear{2013ApJ...770...57B}) and the Rodrigues-Puebla et al. (\citeyear{2017arXiv170304542R}) stellar mass-halo mass relations, respectively. A $V-$band mass-to-light ratio of $M/L_{V} = \textrm{0.3}$ has assumed, appropriate for, e.g., a 1 Gyr old, Z/H=-1 stellar population.}
\label{fig:abundances_young}
\end{figure*}

In Section \ref{results} we present the result of our model, assuming a $V-$band mass-to-light ratio of $M/L_{V} = \textrm{2.0}$, appropriate for old, metal poor stellar populations. 
Here we present the result of our model assuming a lower $V-$band mass-to-light ratio of $M/L_{V} = \textrm{0.3}$, appropriate for younger (1 Gyr), metal poor stellar populations. We use the Flexible Stellar Population Synthesis (FSPS) models to estimate the mass-to-light ratio of 0.3 for the younger population (\citealt{2012ApJ...747...69C}).
The results are presented in Figures \ref{fig:min_stellar_mass_young} and \ref{fig:abundances_young}.

As expected, younger stellar populations results in brighter, more easily detected galaxies and thus increase the expected detection rates of field dwarf galaxies. 
The predicted number of detected field dwarfs in the Local Volume ($3-10 \textrm{ Mpc}$) is now $~1.2$ galaxies per square degree, for limiting surface brightness fainter than $\mu_{\textrm{eff},\textrm{lim}} \sim 29.5 \textrm{ mag arcsec}^{-2}$ and a spatial resolution better than $\theta \sim 3.5 "$, compared to $~0.35$ for a mass-to-light ratio of 2.0, when adopting the B13 stellar mass-halo mass relation (left panels). Similarly to the results presented in Section \ref{results}, adopting the RP17 relation reduces the reduces the number of detected field dwarfs for the same observational limits.

\end{document}